\documentclass[lettersize,journal]{IEEEtran}
\usepackage{amsmath,amsfonts,amssymb}
\usepackage{bm}
\usepackage{algorithm}
\usepackage{algorithmic}
\usepackage{array}
\usepackage{textcomp}
\usepackage{stfloats}
\usepackage{url}
\usepackage{verbatim}
\usepackage{graphicx}
\usepackage{caption}
\usepackage{cite}
\ifCLASSOPTIONcompsoc
  \usepackage[caption=false,font=normalsize,labelfont=sf,textfont=sf]{subfig}
\else
  \usepackage[caption=false,font=footnotesize]{subfig}
\fi

\begin{document}

\title{Learning State-Augmented Policies for \\ Information Routing in Communication Networks}

\author{Sourajit Das, Navid NaderiAlizadeh, Alejandro Ribeiro,~\IEEEmembership{Member,~IEEE}
\thanks{Sourajit Das and Alejandro Ribeiro are with the University of Pennsylvania, Philadelphia, PA 19104 USA (emails: \{dassour, aribeiro\}@seas.upenn.edu).}
\thanks{Navid NaderiAlizadeh is with Duke University, Durham, NC 27705 USA (email: navid.naderi@duke.edu).}}



\maketitle

\begin{abstract}
This paper examines the problem of information routing in a large-scale communication network, which can be formulated as a constrained statistical learning problem having access to only local information. We delineate a novel State Augmentation (SA) strategy to maximize the aggregate information at source nodes using graph neural network (GNN) architectures, by deploying graph convolutions over the topological links of the communication network. The proposed technique leverages only the local information available at each node and efficiently routes desired information to the destination nodes. We leverage an unsupervised learning procedure to convert the output of the GNN architecture to optimal information routing strategies. In the experiments, we perform the evaluation on real-time network topologies to validate our algorithms. Numerical simulations depict the improved performance of the proposed method in training a GNN parameterization as compared to baseline algorithms.
\end{abstract}


\section{Introduction}
\IEEEPARstart{T}{he} modern day telecommunication scenario is experiencing rampant growth of ubiquitous wireless devices and intelligent systems. With the inception of $5^{\text{th}}$ generation (5G) communication networks, there has been a growing need for efficient utilization of available resources and delivery of services. Moreover, the proliferation of artificial intelligence (AI) has been pivotal in addressing the challenging problems of wireless communications in more superior ways. Following its success in areas such as computer vision and natural language processing, machine learning (ML) and AI have taken over the research efforts in wireless networks and have opened new avenues to address problems in which learning-based solutions outperform 
conventional, non-learning-based methods \cite{mao2018deep}. 

Communication networks, including both wired and wireless, present a plethora of complex features and network control targets that have significant impacts on communication performances, such as radio resource management, traffic congestion control, queue management, etc. There are significant works contributing to address such challenges which include stochastic network utility maximization (NUM) \cite{liu2015joint}, Radio Resource Management (RRM) \cite{eisen2020optimal, naderializadeh2023learning, naderializadeh2022state, uslu2024learning}, routing \cite{xia2018utility, zargham2013accelerated, ribeiro2009stochastic}, and scheduling \cite{xia2017stochastic, xia2014distributed, georgiadis2006resource}. They are formulated as constrained optimization problems of a utility function considering stochastic dynamics of user traffic and fading wireless channels.

Mao \textit{et al.} present a comprehensive survey of ML algorithms in intelligent wireless networks to enhance different network functions such as resource allocation, routing layer path search, traffic balancing, sensing data compression, etc \cite{mao2018deep}. With similar approaches, the networking research has found another scope in the form of Reinforcement Learning (RL) to develop efficient ML based solutions for network engineering \cite{bernardez2023magnneto, valadarsky2017learning, xu2018experience, geng2020multi}. Many ML-based approaches that have been proposed in the communication networking area have relied on supervised learning procedures to train a neural network that can mimic the existing system heuristics thereby reducing the computational cost in the execution phase \cite{sun2017learning, lei2017deep,xu2019energy, van2019sum}. Unfortunately such learning methods mandate the need for training sets to build solutions for the system model and may not have the potential to exceed heuristic performances. An alternate approach would be to treat such communication problem as a statistical regression model which solve the optimization problem directly instead of relying on training sets \cite{eisen2020optimal,de2018team,meng2020power,  cui2019spatial}. Such unsupervised learning outperforms the capabilities of supervised learning as it can be applied to any routing or resource allocation problem with a potential to exceed existing heuristics. 

From a network layer viewpoint, the devices in a communication network can be seen as exchanging information packets using communication protocols based on 3 main categories; distance vector, link state and path vector \cite{MEDHI201864}. Information packets are generated by the network nodes and denoted for delivery to assigned destinations. It is crucial that the nodes are also required to determine routes and link schedules to ensure successful delivery of the packets under given traffic conditions. Every node handles packets that are locally generated along with the packets which are received from neighboring nodes. Thus the goal of every node in the network is to determine the next hop of suitable neighbors for each flow conducive to successful routing of the packet to the destination node. 

With the challenge of training network models for modern communication systems, fully connected neural networks (FCNNs) seemed to be a strong contender 
\cite{jia2014caffe, sun2017learning, eisen2020optimal, peng2017modulation}. The scalability of the message signals in time and space has led CNNs to solve network routing problems recently \cite{zhang2023admire, he2020machine}. However, these CNNs do not generalize well enough to large scale scenarios which involve large network graphs with multiple systems parameters \cite{shen2022graph}. Further, they lack the ability to transfer and scale to other networks due to permutation-invariance of the communication networks. Hence, we use Graph Neural Network (GNN) architectures in this paper to mitigate the shortcoming of a Fully Connected Neural Network (FCNN), and accomplish benefits such as scalability, transferability and permutation invariance \cite{henaff2015deep, gama2018convolutional}. The GNNs have another important property which allows them to be implemented in a distributed manner while this property does not apply to the FCNNs or CNNs \cite{solodova2024graph}.

Similar to~\cite{eisen2020optimal, ribeiro2012optimal, naderializadeh2023learning, naderializadeh2022state, xia2018utility, zargham2013accelerated, ribeiro2009stochastic, xia2017stochastic, xia2014distributed}, we consider the problem of network utility maximization subject to multiple constraints. The utility function and some of the constraints rely upon the average performance of the nodes across the network while one of the constraints is an instantaneous constraint. The objective focuses on improving the routing and scheduling in packet based networks. The routing decisions made at the node level are crucial to ensure the network stability and thereby forcing the queue lengths to be stabilized in the steady state. In pursuit of the above network optimization, we encounter certain challenges or constraints which need to be ensured for the systems to be feasible. Generally, one solves such problems by switching to the Lagrangian dual domain, where a single objective function is maximized over the primal variables and minimized over the dual variables. The primal variable correspond to the objective function while the dual variable designate the constraints in the original routing problem. Although such primal-dual methods help to reach optimal solutions, the null duality gap persists even with near-universal parameterizations, such as fully connected neural networks (FCNN) \cite{naderializadeh2022state}. Moreover, it is uncertain whether they lead to routing decisions which satisfy the constraints in the original optimization problem while lacking feasibility guarantees. 

In this paper, we take an alternate approach of state augmentation based constrained learning. More specifically, we take advantage of the fact that dual variables in constrained optimization provide the degree of constraints' violation or satisfaction over a period of time \cite{calvo2021state}. We leverage the above concept to augment the typical communication network state with dual variables at each time instant, to be used as dynamic inputs to the routing policy. The above approach of incorporating dual variables to the routing policy enables us to train the policy to adapt its decision to instantaneous channel states as well as ensuring adaptation to constraint satisfaction. 

In summary, we make the following contributions in this paper:
\begin{itemize}
    \item We use Graph Neural Networks (GNNs) to learn efficient routing strategies in decentralized communication networks of varying sizes.
    \item We leverage the Method of Multipliers (MoM) to achieve faster convergence than conventional dual descent methods.
    \item We utilize State Augmentation methods to learn near-optimal solutions with a finite number of iterations.
\end{itemize}

The rest of the paper is organized as follows. We start with some of the related works pertaining to learning and optimization in communication networks in Section \ref{sec:rel_work}. We present the problem formulation as a constrained network utility maximization problem in Section \ref{sec:problem}. We discuss some of the standard optimization techniques in Sections \ref{sec:dual_desc}, \ref{sec:mom} and \ref{sec:admm}. 
The parameterization of the above learning algorithm using Graph Neural Networks is presented in Section \ref{sec:gnn_par}. Furthermore, we discuss the fundamental principles of augmented Lagrangian and GNN Learning to understand their application in our problem in Section \ref{sec:gnn_mom}. In Section \ref{sec:state_aug}, we describe our proposed state-augmented algorithm to optimize routing strategies for the network optimization problem. The performance of the different approaches are discussed and compared in the results of Section \ref{sec:results}. Finally, we conclude the paper in Section \ref{sec:conclusion}.

\section{Related Work} \label{sec:rel_work}

The progress in communication networks primarily consisted of solutions to solving the problem of resource management. In \cite{uslu2024learning, naderializadeh2023learning, naderializadeh2022state}, the authors addresses the challenge of optimizing the power allocation in a wireless communication system using GNNs. \cite{eisen2019learning} deals with design of optimal resource allocation policies using Deep Neural Networks (DNNs). In \cite{nasir2019multi, liang2019deep}, the authors use Deep Reinforcement Learning (DRL) techniques for transmit power control in wireless networks to mitigate interference and network efficiency.  In terms of routing, there has been few works which aim to improve network capacity for communication networks. In \cite{liu2006cross}, the authors provide a mathematical solution comprising of gradient based methods to jointly optimize power and multi-path routing. \cite{xia2018utility} addresses the challenges in stochastic NUM under heavy-tailed conditions
and proposed a stochastic gradient descent routing algorithm to achieve optimality and stability. Paper \cite{ribeiro2009stochastic} introduces a generalization of the backpressure algorithm which randomizes routing and scheduling in ad-hoc networks to enhance delay performance. The authors in \cite{zargham2013accelerated} enhance the backpressure algorithm further by using improvised dual descent methods for faster convergence and stability in dynamic network conditions. We derive the motivation for addressing the routing issues majorly from \cite{zargham2013accelerated, ribeiro2009stochastic}. Here we optimize our routing and scheduling policies via GNN parameterization to imitate the Method of Multipliers (MoM) which produces faster convergence without sacrificing optimality. The closest prior work to the current study is \cite{naderializadeh2022state} where we echo the authors' approach to use state augmentation methods to achieve faster convergence and near-optimal solutions with a finite number of iterations, while extending it beyond conventional primal-dual algorithms.





\section{Problem Statement} \label{sec:problem}
We consider a communication network, modeled as a graph $\mathcal{G} = (\mathcal{V, E})$ where $\mathcal{V}$ is the set of nodes and $\mathcal{E} \subseteq \mathcal{V} \times \mathcal{V}$ is the set of links between nodes. We denote the capacity of a link $(i,j) \in \mathcal{E}$ as $C_{ij}$ and we define the neighborhood of $i$ as the set $\mathcal{N}_i = \{j \in \mathcal{V} | (i,j) \in \mathcal{E} \}$ of nodes $j$ that can communicate directly to $i$. The nodes communicate with each other by exchanging information packets in different flows. We use $\mathcal{K}$ to denote the set of information flows with the destination of flow $k \in \mathcal{K}$ being the node $o_k \in \mathcal{V}$.

At a given time instant $t$, node $i\neq o_k $ generates a random number of packets, denoted by $A_i^k(t)$, which is to be delivered to the destination $o_k$. We assume that the random variable(s) $A_i^k(t)$ are independent and identically distributed (\textit{iid}) across time with the expected value $\mathbb{E}[A_i^k(t)]=A_i^k$. Within the same time instant, node $i$ routes $r_{ij}^k(t)$ units of information packets to every neighboring node $j \in \mathcal{N}_i$ and also receives $r_{ji}^k(t)$ packets from neighboring node $j$. The difference between the total number of received packets $A_i^k(t) + \sum_{j \in \mathcal{N}_i} r_{ji}^k(t)$ and the sum of total number of transmitted packets $\sum_{j \in \mathcal{N}_i} r_{ij}^k(t)$ is added to the local queue or subtracted if the resultant quantity is negative. Hence, we can iteratively define the queue length of the packets of flow $k$ at node $i$, which we denote as $q_i^k(t)$, according to the following equation,
\begin{equation} \label{queueupdate}
    q_i^k(t+1) = \Bigr[q_i^k(t) + A_i^k(t) + \sum_{j \in \mathcal{N}} r_{ji}^k(t) - r_{ij}^k(t) \Bigr]^+,  
\end{equation}
where we add a projection on the non-negative orthant to account for the fact that the number of packets in the queue will always be non-negative. Note that \eqref{queueupdate} is mentioned for all nodes $i\neq o_k $ as packets routed to their destination will be discarded from the network.

\subsection{Problem Formulation}
\label{ssec:problem_subhead}

Now, we consider the above communication network operating over a series of time steps $t \in \{0, 1, 2,..,T-1 \}$, where at each time step $t$, the set of channel capacities, or the \textit{network state}, is denoted as $\mathbf{C}_t \in \mathcal{C}$. For the above network state, let $\mathbf{p}(\mathbf{C}_t)$ denote the vector for routing decisions across the network, where $\mathbf{p}:\mathcal{C} \rightarrow \mathbb{R}^{n \times n \times F}$ represents the routing function. These routing decisions eventually lead to the network performance vector $\mathbf{f}(\mathbf{C}_t, \mathbf{p}(\mathbf{C}_t)) \in \mathbb{R}^b$, where $\mathbf{f}:\mathcal{C} \times \mathbb{R}^{n \times n \times F} \rightarrow \mathbb{R}^b$ represents the performance function.

Furthermore we consider an auxiliary optimization variable $a_i^k(t) \geq A_i^k(t)$ to ensure we generate as many packets at node $i$ for flow $k$ as possible. In real-time, $A_i^k(t)$ is the actual number of packets present in the system, which is used to update the queue length defined in \eqref{queueupdate}. As a general approach in \cite{naderializadeh2022state}, we start with a concave utility function $\mathcal{U}: \mathbb{R}^b \rightarrow \mathbb{R}$ and a set of constraints $\mathbf{g}: \mathbb{R}^b \rightarrow \mathbb{R}^c$. The generic routing problem can be defined as below.
\begin{subequations} \label{eq:concave_util}
\begin{align}
    \max_{[\mathbf{p}(\mathbf{C}_t)]_{t=0}^{T-1}} \hspace{0.5cm} \mathcal{U} \Biggl(\frac{1}{T} \sum_{t=0}^{T-1} \mathbf{f}(\mathbf{C}_t, \mathbf{p}(\mathbf{C}_t))\Biggl) \\
    s.t. \hspace{1cm} \mathbf{g}\Biggl(\frac{1}{T} \sum_{t=0}^{T-1} \mathbf{f}(\mathbf{C}_t, \mathbf{p}(\mathbf{C}_t)) \Biggl) & \geq 0,
\end{align}
\end{subequations}
where the objective function and the constraints are derived based on the \textit{ergodic average} of the network performance \textit{i.e.} $\frac{1}{T} \sum_{t=0}^{T-1} \mathbf{f}(\mathbf{C}_t, \mathbf{p}(\mathbf{C}_t))$. Therefore the aim of the routing algorithm is to determine the optimal vector of routing decisions $\mathbf{p}(\mathbf{C}_t)$ for any given network state $\mathbf{C}_t \in \mathcal{C}$. 

For the routing problem under consideration, we start with the utility function which considers maximizing the information packets at all nodes $i$ and over all flows $k$. The concave objective function can be written as below.
\begin{equation} \label{eq:util_function}
    \mathcal{U} \Biggl(\frac{1}{T} \sum_{t=0}^{T-1} \mathbf{f}(\mathbf{C}_t, \mathbf{p}(\mathbf{C}_t))\Biggl) = \sum_{k \in \mathcal{K}} \sum_{i \in \mathcal{V}} \log \Biggl(\frac{1}{T} \sum_{t=0}^{T-1} a_i^k(t) \Biggl).
\end{equation}

We consider three sets of constraints with the given network state to optimize the above objective function. The first set of constraints can be defined as the \textit{routing constraints}, which state that the sum of the total number of received packets, $r_{ji}^k(t)$ and the local packets, $a_i^k(t)$ at node $i$ should be equal or less than the total number of packets transmitted, $r_{ij}^k(t)$ from node $i$. 
\begin{equation} \label{eq:flow_constr}
    a_i^k(t) + \sum_{j \in \mathcal{N}_i} r_{ji}^k(t) \leq \sum_{j \in \mathcal{N}_i} r_{ij}^k(t).
\end{equation}
The second set of constraints are the \textit{minimum constraints} which assume that the node $i$ has a minimum number of local packets to ensure reliable transmission to other nodes.
\begin{equation} \label{eq:cap_constr}
    a_i^k(t) \geq A_i^k(t).
\end{equation}
The third set of constraints can be termed as the \textit{capacity constraints}, as they define the maximum capacity of a channel between two nodes, \textit{i.e.} the sum of the total number of packets over all the flows in a link between node $i$ and node $j$ can \emph{not} be more than the maximum capacity of the channel.
\begin{equation} \label{eq:min_constr}
    \sum_{k \in K} r_{ij}^k(t) \leq C_{ij}.
\end{equation}

Plugging 
\eqref{eq:util_function}-\eqref{eq:min_constr} into the generic formulation in~\eqref{eq:concave_util}, the routing optimization problem can be written as,
\begin{subequations} \label{eq:prob_state}
\begin{align}
    \max_{[a_i^k(t), r_{ij}^k(t)]_{t=0}^{T-1}} \hspace{0.25cm}  \sum_{k \in \mathcal{K}} \sum_{i \in \mathcal{V}} \log \Biggl(\frac{1}{T} \sum_{t=0}^{T-1} a_i^k(t)\Biggl) \label{eq:7a} \\
    \sum_{j \in \mathcal{N}_i} r_{ij}^k(t) - \sum_{j \in \mathcal{N}_i} r_{ji}^k(t) - a_i^k(t) \geq 0 \label{eq:7b} \\
    a_i^k(t) - A_i^k(t) \geq 0 \label{eq:7c} \\
    C_{ij}(t) - \sum_{k \in K} r_{ij}^k(t) \geq 0 \label{eq:7d} 
\end{align}
\end{subequations}
\section{Solving for Gradient Based Routing in the Lagrangian Domain} \label{sec:dual_desc}
Since the optimization problem in \eqref{eq:prob_state} involves a concave utility function, we can use a gradient based dual descent algorithm. We introduce a dual variable $\bm{\mu} \in \mathbb{R}_+^c$, also called as the Lagrangian multiplier corresponding to the constraint \eqref{eq:7b}. We keep the constraints \eqref{eq:7c} and \eqref{eq:7d} implicit while considering $T=1$ for convenience in analysis. Now we can express the Lagrangian as follows.
\begin{multline} \label{eq:dd_lagangian}
    \mathcal{L} (\bm{a, r, \mu}) = \sum_i \sum_k \log ( a_i^k ) \\
    + \sum_i \sum_k \mu_i^k \Bigg( \sum_{j \in \mathcal{N}_i} r_{ij}^k - \sum_{j \in \mathcal{N}_i} r_{ji}^k - a_i^k\Bigg)
\end{multline}
The Lagrangian in \eqref{eq:dd_lagangian} can be maximized using a standardised gradient based algorithm. More specifically, a dual descent like the primal-dual learning works to maximize $\mathcal{L}$ over $\bm{a}$ and $\bm{r}$ while minimizing over the dual variables $\bm{\mu}$ at the same time.
\begin{equation} \label{eq:dd_lag_update}
    \mathcal{L}^* = \min_{\bm{\mu}} \max_{\bm{a,r}} \hspace{0.25cm} \mathcal{L}(\bm{a}, \bm{r}, \bm{\mu})
\end{equation}
For a given iteration index $m \in \{1,2,..,M\}$, the primal variables $a_i^k$ and $r_{ij}^k$ can be updated according to the following equations \eqref{dd:a_update} and \eqref{dd:r_update},
\begin{equation} \label{dd:a_update}
    [{a_i^k}]_{m+1} = [{a_i^k}]_{m} + \eta_{\bm{\phi}} \bm{\nabla}_{\phi} \mathcal{L} (\bm{a, r, \mu})
\end{equation}
\begin{equation} \label{dd:r_update}
    [{r_{ij}^k}]_{m+1} = [{r_{ij}^k}]_m + \eta_{\bm{\phi}} \bm{\nabla}_{\phi} \mathcal{L} (\bm{a, r, \mu})
\end{equation}
Now the gradient descent on the dual variable can be expressed as $\bm{\mu}_{m+1} =  \bm{\mu}_m - \eta_{\bm{\mu}} \bm{\nabla}_{\phi} \mathcal{L} (\bm{a, r, \mu})$. Since the Lagrangian is linear in the dual variable, the gradient is easy to compute and $\bm{\mu}$ can be updated recursively as 
\begin{multline} \label{dd:dual_update}
    \bm{\mu}_{m+1} = \Bigg[ \bm{\mu}_m - \eta_{\bm{\mu}} \Big( \sum_{j \in \mathcal{N}_i} r_{ij}^k - \sum_{j \in \mathcal{N}_i} r_{ji}^k - a_i^k\Big) \Bigg]^+
\end{multline}
We consider $[.]^+$ to account for the non-negative orthant and can be defined as $[x]^+ = max(x,0)$. Moreover, $\eta_{\bm{\mu}}$ and $\eta_{\bm{\phi}}$ denote the corresponding learning rates for the primal variables ($\bm{a,r}$) and dual variable ($\bm{\mu}$). The above routing optimization problem can deliver near-optimal and feasible results when run for a large enough number of time steps. Standard gradient descent techniques such as primal-dual learning may not guarantee a feasible set for routing decisions, whereas such a feasibility can be guaranteed for the routing algorithms in \eqref{dd:a_update}, \eqref{dd:r_update}, \eqref{dd:dual_update}. Even though the above algorithm generates optimal solutions with feasibility, a major drawback of the above dual descent algorithm can be attributed to its slow convergence rate. As an alternative approach, we opt for the Method of Multipliers which we discuss in the upcoming section.

\section{Solution using the Augmented Lagrangian: Method of Multipliers} \label{sec:mom}
When we consider an optimization problem to be solved in the Lagrangian domain, dual descent method serves as a good starting point to reach optimal solution. Unfortunately, such an approach is known to have proven disadvantages, the most notable being its extremely slow convergence rate and the need for strictly convex objective functions. Hence the Augmented Lagrangian or the Method of Multipliers (MoM) serves as an excellent alternate to alleviate the above problems of dual descent methods \cite{chatzipanagiotis2015augmented, bertsekas2015parallel, ruszczynski2011nonlinear}. To start with the algorithm, we take care of the inequality constraint \eqref{eq:7a} by converting it into an equality constraint as shown below,
\begin{equation} \label{eq:ineq_to_equal}
    \sum_{j \in \mathcal{N}_i} r_{ij}^k(t) - \sum_{j \in \mathcal{N}_i} r_{ji}^k(t) - a_i^k(t) - z_i^k(t) = 0
\end{equation}
where $z_i^k \geq 0$ is an auxiliary variable. 
The Augmented Lagrangian for the optimization problem in \eqref{eq:prob_state} can now be written as
\begin{multline} \label{aug_lag_mom}
     \mathcal{L}_{\rho} (\bm{a, r, z, \mu}) = \sum_{k\in\mathcal{K}} \sum_{i\in\mathcal{V}} \frac{1}{T} \sum_{t=0}^{T-1} \log ( a_i^k ) \\
    + \sum_{k\in\mathcal{K}}\sum_{i\in\mathcal{V}} \frac{1}{T} \sum_{t=0}^{T-1} \Bigg\{ \mu_i^k \Bigg( \sum_{j \in \mathcal{N}_i} r_{ij}^k - \sum_{j \in \mathcal{N}_i} r_{ji}^k - a_i^k - z_i^k \Bigg) \\
    + \frac{\rho}{2} \Biggl|\!\Biggl| \sum_{j \in \mathcal{N}_i} r_{ij}^k - \sum_{j \in \mathcal{N}_i} r_{ji}^k - a_i^k - z_i^k \Biggr|\!\Biggr|^2 \Bigg\}
\end{multline}
where $\rho > 0$ is a penalty parameter and $\bm{\mu} \in \mathbb{R}_+^{n \times F}$ is the dual multiplier associated with the constraint \eqref{eq:7b}. We keep the constraints \eqref{eq:7c} and \eqref{eq:7d} implicit, i.e., we force our solutions to the optimization problem to automatically satisfy them. The penalty method solves the above optimization problem for various values of $\bm{\mu}$ and $\rho$.
\begin{subequations} \label{eq:MoM_algo}
\begin{align}
    \bm{a}^{m+1}, \bm{r}^{m+1}, \bm{z}^{m+1} = \arg \max_{\bm{a}, \bm{r}, \bm{z}} \mathcal{L}_{\rho} (\bm{a}, \bm{r}, \bm{z}, \bm{\mu}^m) \\
    \bm{\mu}^{m+1} = \arg \min_{\bm{\nu}} \mathcal{L}_{\rho} (\bm{a}^{m+1}, \bm{r}^{m+1}, \bm{z}^{m+1}, \bm{\mu}^m).\label{eq:mom_min_dual}
    \end{align}
\end{subequations}
Note that since $\mathcal{L}_{\rho} (\bm{a, r, z}, \bm{\mu})$ is a linear function of the dual variables $\bm{\mu}$, the minimization in~\eqref{eq:mom_min_dual} can be implemented using gradient descent, i.e.,
\begin{equation} \label{eq:mom_dual}
    (\mu_i^k)^{m+1} = \Bigg[(\mu_i^k)^{m} - \rho^m \bigg(\sum_{j \in \mathcal{N}_i} r_{ij}^k - \sum_{j \in \mathcal{N}_i} r_{ji}^k - (a_i^k)^m \bigg) \Bigg]^+. 
\end{equation}

The convergence of the MoM algorithm is ensured under certain conditions (in particular, when the maximization over $\bm{z}$ 
has an optimal solution independent of the initialization) \cite{chatzipanagiotis2017convergence}. However, a critical disadvantage of the above method is that MoM loses decomposability, thereby making it inefficient in complex distributed systems. This motivates the use of 
another variation of the above process called \textit{The Alternating Direction Method of Multipliers}, which we discuss in the next section.

\section{The Alternating Direction Method of Multipliers (ADMM)} \label{sec:admm}
The Alternating Direction Method of Multipliers or ADMM combines the best of both Dual Descent and Method of Multipliers algorithms. The ADMM exploits the decomposable structure of the augmented Lagrangian. Basically ADMM solves the original convex optimization problem by breaking it into smaller sub-problems that can be solved efficiently \cite{su2022distributed, chatzipanagiotis2017convergence, bertsekas2015parallel}. The augmented Lagrangian for our optimization problem in \eqref{eq:prob_state} is given by 
\begin{multline} \label{aug_lag_admm1}
    \mathcal{L}_{\rho} (\bm{a}, \bm{r}, \bm{z}, \bm{\nu}) = \sum_i \sum_k \log ( a_i^k ) + g(\bm{z})\\
    + \sum_i \sum_k \Bigg\{ \nu_i^k \Bigg( \sum_{j \in \mathcal{N}_i} r_{ij}^k - \sum_{j \in \mathcal{N}_i} r_{ji}^k - a_i^k - z_i^k \Bigg) \\
    - \frac{\rho}{2} \Bigg| \sum_{j \in \mathcal{N}_i} r_{ij}^k - \sum_{j \in \mathcal{N}_i} r_{ji}^k - a_i^k - z_i^k \Bigg|^2 \Bigg\}
\end{multline}
where $g$ is an indicator function on $\bm{z}$ defined by
\begin{equation} \label{eq:indicator}
    g(\bm{z}) = 
    \begin{cases}
    0, & \text{if } \bm{z} \geq 0 \\
    - \infty, & \text{otherwise}
    \end{cases}
\end{equation}
and $\nu$ is the dual variable associated with the constraint while $\rho$ is the penalty parameter. The general ADMM iteration follows,
\begin{subequations}
    \begin{align}
        \bm{a}^{m+1}, \bm{r}^{m+1} = \arg \max_{\bm{a}, \bm{r}} \mathcal{L}_{\rho} (\bm{a}, \bm{r}, \bm{z}^m, \bm{\nu}^m) \\
        \bm{z}^{m+1} = \arg \max_{\bm{z}} \mathcal{L}_{\rho} (\bm{a}^{m+1}, \bm{r}^{m+1}, \bm{z}, \bm{\nu}^m) \\
        \bm{\nu}^{m+1} = \arg \min_{\bm{\nu}} \mathcal{L}_{\rho} (\bm{a}^{m+1}, \bm{r}^{m+1}, \bm{z}^{m+1}, \bm{\nu}^m)
    \end{align}
\end{subequations}
The prime difference between the MoM and ADMM is that the primal maximization here is split into two parts instead of optimizing over ($\bm{a},\bm{r},\bm{z}$) jointly \cite{boyd2011distributed}. 
Alternatively, we can also use a scaled form of the above algorithm by considering $\mu = \frac{1}{\rho}\nu$. Substituting this in the above augmented Lagrangian, we get, 

\begin{multline} \label{aug_lag_admm2}
    \mathcal{L}_{\rho} (\bm{a}, \bm{r}, \bm{z}, \bm{\mu}) = \sum_i \sum_k \log ( a_i^k ) + g(\bm{z})\\
    - \frac{\rho}{2} \Bigg|\Bigg| \sum_{j \in \mathcal{N}_i} r_{ij}^k - \sum_{j \in \mathcal{N}_i} r_{ji}^k - a_i^k - z_i^k - \mu_i^k \Bigg|\Bigg|^2 - \frac{\rho}{2} \Big|\Big| \mu_i^k \Big|\Big|^2
\end{multline}

Now the steps of ADMM can be written as,

\begin{subequations} \label{eq:admm_scaled}    
    \begin{align} \label{admm_scaled_a} 
    \begin{split}
        \bm{a}^{m+1}, \bm{r}^{m+1} = \arg \max_{\bm{a}, \bm{r}} \Bigg\{ \sum_i \sum_k \log ( a_i^k ) - \frac{\rho}{2} \Bigg|\Bigg| \sum_{j \in \mathcal{N}_i} r_{ij}^k - \\
        \sum_{j \in \mathcal{N}_i} r_{ji}^k - a_i^k - [z_i^k]^m - [\mu_i^k]^m \Bigg|\Bigg|^2 - \frac{\rho}{2} \Big|\Big| [\mu_i^k]^m \Big|\Big|^2 \Bigg\} \\
    \end{split}\\    
    \begin{split} \label{admm_scaled_b} 
        \bm{z}^{m+1} = \arg \max_{\bm{z}} \Bigg\{g(\bm{z}) - \frac{\rho}{2} \Bigg|\Bigg| \sum_{j \in \mathcal{N}_i} [r_{ij}^k]^{m+1} - \\
        \sum_{j \in \mathcal{N}_i} [r_{ji}^k]^{m+1} - [a_i^k]^{m+1} - z_i^k - \mu_i^k \Bigg|\Bigg|^2 - \frac{\rho}{2} \Big|\Big| \mu_i^k \Big|\Big|^2\Bigg\} \\
    \end{split}\\
    \begin{split} \label{admm_scaled_c} 
        [\mu_i^k]^{m+1} = \Bigg[ [\mu_i^k]^m - \Bigg( \sum_{j \in \mathcal{N}_i} [r_{ij}^k]^{m+1} \\
        - \sum_{j \in \mathcal{N}_i} [r_{ji}^k]^{m+1} - [a_i^k]^{m+1} - [z_i^k]^{m+1} \Bigg) \Bigg]^+
        \end{split}
    \end{align}
\end{subequations} 

Here $\bm{\mu}$ is the scaled dual variable for the constraint considered in the optimization problem. Just like the name suggests, note that the iterative optimization process is carried out via alternating the direction of the variables by sequentially solving the subproblems \eqref{admm_scaled_a} and \eqref{admm_scaled_b}. Since the optimal solution is obtained in each subproblem, we can conclude that the augmented Lagrangian is maximized after sufficient number of iterations. We select a suitable penalty parameter $0 < \rho < 1$ to ensure the convergence of ADMM but larger choices may generate improved results for a very small number of iterations. We observe that the unparameterized form of ADMM works effectively for consensus problems which have a decomposable structure but real-time problems often require more general framework to handle wider range of scenarios as in case of the Method of Multipliers (MoM). Even though MoM is faster in terms of convergence than ADMM, unfortunately it is a centralized algorithm unlike the ADMM. Thus we propose to use Graph Neural Networks (GNNs) that learn to mimic the solution of MoM and implement a decentralized system which can potentially outperform the decentralized ADMM solution.
 
\section{Graph Neural Networks based Parameterization for Routing Optimization} \label{sec:gnn_par}
We observe that \eqref{eq:prob_state} is an infinite-dimensional optimization problem because we need to find $a_i^k$ and $r_{ij}^k$ for any given channel state $\mathbf{C}_t$ and input $A_i^k(t)$ and these are hard to obtain in practice. Alternatively we use a parameterized model which takes $\mathbf{C}_t$ and $A_i^k(t)$ as input and generates the desired decisions $a_i^k(t)$ and $r_{ij}^k(t)$ as output. We take advantage of Graph Neural Networks (GNNs) for parameterizing the information routing policy. These architectures are a family of neural networks specially designed to operate over graph structured data \cite{battaglia2018relational, wu2020comprehensive}. Similarly to the recent studies in~\cite{eisen2020optimal, naderializadeh2023learning, naderializadeh2022state, lee2020graph, shen2019graph}, we use GNNs as they have been seen to provide multiple benefits such as permutation equivariance, scalability and transferability to other networks. Recall that the network in our considered problem can represented as a graph $\mathcal{G} = (\mathcal{V}, \mathcal{E}, \mathbf{z}_t, w_t)$ at a given time step $t$ where: 
\begin{enumerate}
    \item $\mathcal{V}=[1,2,...,n]$ are the set of graph nodes with each node representing a communicating code in the entire network
    \item $\mathcal{E} \subseteq \mathcal{V} \times \mathcal{V}$ represent the set of directed edges in the graph
    \item $\mathbf{z}_t$ represent the initial node features
    \item $w_t: \mathcal{E} \rightarrow \mathbb{R}$ represents the mapping of each edge to its weight at time $t$. We defined the weight between a pair of nodes to be the normalized channel capacity for our optimization problem, i.e. $w_{ij}(t) = C_{ij}(t)$.
\end{enumerate}
Consider a single time instant $t$, and a slight abuse of notation, we use the channel capacity matrix $\mathbf{C}_t \in \mathbf{R}^{n \times n}$ as an adjacency matrix representation of a graph which links node $i$ to node $j$. The initial node feature $\mathbf{z}_t$ can be reinterpreted as a signal supported on the nodes $i=1,2,..,n$. The GNNs are basically graph convolutional filters supported on the graph $\mathbf{C}_t$ to process the input signal $\mathbf{z}_t \in \mathbf{R}^n$. Since we have a filter, let $\mathbf{h} := \{h_0,...,h_{K-1}\}$ be the set of $K$ filter coefficients. We define $\mathbf{\phi(C_t)}$ to be the graph filter, which is essentially a polynomial on the graph representation applied linearly to the input signal $\mathbf{z}_t$ \cite{sandryhaila2014big}. The resulting output signal of the convolution operation can be given as,
\begin{equation} \label{outputy}
    \mathbf{y}_t = \mathbf{\phi(C_t)z_t} = \sum_{k=0}^{K-1}h_k \mathbf{C_t}^k \mathbf{z_t}.
\end{equation}
Note that the filter $\mathbf{\phi(C_t)}$ has the characteristics of a linear shift invariant filter and the graph $\mathbf{C_t}$ is called as a Graph Shift Operator (GSO) \cite{sandryhaila2014big}.
The GNN processes the input node and edge features using $L$ layers, where the $l^{th}$ layer transforms the node features at $l-1$ layer, i.e. $\mathbf{Y}_t^{l-1} \in \mathbb{R}^{m \times F_{l-1}}$, to the node features at layer l, $\mathbf{Y}_t^l \in \mathbb{R}^{m \times F_l}$. The filter in the $l^{th}$ can be applied to the output of the $l-1$ layer to produce the feature $\mathbf{y}_t^{l}$ which can be written as
\begin{equation} \label{outputyl}
    \mathbf{y}_t^l = \mathbf{\phi}^l\mathbf{(C_t)z_t}^{l-1} = \sum_{k=0}^{K_l-1}h_{lk} \mathbf{C}_t^k \mathbf{z}_t^{l-1}.
\end{equation}
The intermediate features are then passed through a pointwise non-linearity function to generate the output of the $l^{th}$ layer, i.e.
\begin{equation} \label{outputztl}
    \mathbf{z}_t^l = \mathbf{\sigma}[\mathbf{y}_t^l] = \mathbf{\sigma} \Bigg[ \sum_{k=0}^{K_l-1}h_{lk} \mathbf{C}_t^k \mathbf{z}_t^{l-1} \Bigg]
\end{equation}
Finally the GNN is used as a recursive application of the convolution operation in \eqref{outputztl}. It is to be noted that the non-linearity in \eqref{outputztl} is applied to the individual component in each layer and some common choices for $\mathbf{\sigma}$ are rectified linear units (ReLU), absolute value or sigmoid functions \cite{gama2018convolutional, henaff2015deep}. 

The previous expressions for GNNs considered a single graph filter but we can increase the expressive power of the GNN network by using a bank of $F_l$ graph filters \cite{eisen2020optimal}. In other words, these set of filters create multiple features per layer, each of which is processed with a separate graph filter bank. Let's take the output of layer $l-1$ which consists of $F_{l-1}$ features and these features become inputs for the layer $l$, each of which can be processed by the $F_l$ filters $\phi_l^{fg}(\mathbf{C_t})$. The $l^{th}$ layer intermediate feature upon application of these filters can now be given as 
\begin{equation} \label{outputyflg}
    \mathbf{y}_{t,fg}^l = \phi_l^{fg}(\mathbf{C_t}) \mathbf{z}_{t,f}^{l-1} = \sum_{k=0}^{K_l-1}h_{lk}^{fg} \mathbf{C}_t^k \mathbf{z}_{t,f}^{l-1}
\end{equation}
Thus \eqref{outputyflg} shows that the layer $l$ generates $F_{l-1} \times F_l$ intermediate feature $\mathbf{y}_{t,fg}^l$. These features can grow exponentially unless all the features $\mathbf{y}_{t,fg}^l$ for a given value of $g$, are linearly aggregated and passed though a pointwise non-linearity function $\mathbf{\sigma}$ to generate the layer $l$ output. Thus the output $\mathbf{z}_t^l$ at the $l^{th}$ layer can be given as,
\begin{equation}\label{finalzl}
    \mathbf{z}_t^l = \mathbf{\sigma}_l \Bigg[\sum_{f=1}^{F_l} \mathbf{y}_{t,fg}^l \Bigg] = \mathbf{\sigma}_l \Bigg[\sum_{f=1}^{F_l} \phi_l^{fg}(\mathbf{C_t}) \mathbf{z}_{t,f}^{l-1} \Bigg]
\end{equation}
We use \eqref{finalzl} in recurrence to obtain the GNNs for our experiments. The filter coefficients can be grouped in the filter tensor $\phi = [h_{lk}^{fg}]_{l,k,f,g}$ and we define the GNN operator as,
\begin{equation}\label{phiout}
   \mathbf{\Psi}(\mathbf{C}_t,\mathbf{x}_t;\phi) = \mathbf{z}_t^L
\end{equation}
where $\mathbf{x} = \mathbf{z}_t^0$ is the input to the GNN at layer, $l=1$. We consider different number of input features, $F_0$ based on our algorithms described in the later sections. 
Hence the output layer provides the GNN result which can be given as
\begin{equation} \label{r_out}
    \mathbf{y_{out}} = \mathbf{Y}_t^L \in \mathbb{R}^{n \times F_L}
\end{equation}
The resulting output of \eqref{r_out} provides a column vector which can then used along with an intermediate matrix $w_r \in \mathbb{R}^{F_L \times F_L}$ to obtain the routing decisions $r_{ij}^k(t)$. We make sure to pass the resulting matrix through a Softmax filter to ensure the entries of the routing decision matrix are between 0 and 1, i.e.

\begin{equation} \label{p_softmax}
    \mathbf{p}(\mathbf{C}_t,\mathbf{x}_t;\bm{\phi}) = \text{Softmax} (\mathbf{y_{out}} ~ w_r ~ \mathbf{y_{out}}^T),
\end{equation}
where $\text{Softmax}(\mathbf{x})_i = \frac{\exp(x_i)}{\sum_{j=1}^{K}\exp(x_j)} $ for $i=1,2,\dots,K$ and $\mathbf{x}=(x_1,x_2,\dots,x_K) \in \mathbb{R}^K$. The intermediate matrix $w_r$ is a square matrix of shape ($F_L \times F_L$) which helps in generating the resultant routing matrix to be a ($n \times n$) matrix. Thus the above formulation to compute $r_{ij}^k(t)$ takes care of meeting the capacity constraints in \eqref{eq:7d}. 
Similarly we derive the target packets $a_i^k$ by passing the output of the GNN through another linear layer where it is multiplied with a column vector $w_a \in \mathbb{R}^{F_L}$,
\begin{equation} \label{a_ik}
    a_i^k(t) = \big[A_i^k(t) + \mathbf{y_{out}} ~ w_a \big]^+
\end{equation}
We pass the $a_i^k(t)$ through a ReLU filter to account for the fact that packets generated at the nodes are non-negative. The formulation in \eqref{a_ik} ensures that the minimum constraint is satisfied during the optimization process. In summary, the above ways of implementing \eqref{p_softmax} and \eqref{a_ik} satisfy constraints \eqref{eq:7d} and \eqref{eq:7c} which we considered as implicit during the formulation of the Lagrangian. One key take away from the GNN implementation is that with a filter length of $K_l$ at the $l^{th}$ layer, the total number of parameters in GNN is hardly $\sum_{l=1}^{L}K_l$, which is significantly smaller than a fully connected neural network.

\section{GNN Parameterization with Augmented Lagrangian: Method of Multipliers} \label{sec:gnn_mom}
In order to solve the optimization problem in \eqref{eq:prob_state} involving a concave utility function, we use the Method of Multipliers (MoM) to solve for the optimal solution using the GNN parameterization we discussed above. We use $\mathbf{z}_t = A_i^k(t)$ as the input feature to the GNN in order to find the optimal routing policy $\mathbf{p}(\mathbf{C}_t,\mathbf{x}_t;\bm{\phi})$. Using the GNN parameterization in \eqref{phiout}, \eqref{r_out}, \eqref{p_softmax}, we can rewrite the parameterized optimization problem in \eqref{eq:prob_state} as,
\begin{subequations} \label{eq:gnn_par}
\begin{align}
    \max_{\phi, w_r, w_a} \hspace{0.5cm} \mathcal{U} \Biggl(\frac{1}{T} \sum_{t=0}^{T-1} \mathbf{f}(\mathbf{C}_t, \mathbf{p}(\mathbf{C}_t;\bm{\phi}))\Biggl) \\
    s.t. \hspace{1cm} \mathbf{g}\Biggl(\frac{1}{T} \sum_{t=0}^{T-1} \mathbf{f}(\mathbf{C}_t, \mathbf{p}(\mathbf{C}_t;\bm{\phi})) \Biggl) & \geq 0 \label{gnnconst1}  
\end{align}
\end{subequations}
The Augmented Lagrangian for the problem in \eqref{eq:gnn_par} is given by
\begin{multline} \label{aug_lag1}
    \mathcal{L}(\bm{\phi}, \bm{\mu}) = \mathcal{U} \Biggl(\frac{1}{T} \sum_{t=0}^{T-1} \mathbf{f}(\mathbf{C}_t, \mathbf{p}(\mathbf{C}_t;\bm{\phi}))\Biggl) \\ 
     + \bm{\mu}^T  \mathbf{g}\Biggl(\frac{1}{T} \sum_{t=0}^{T-1} \mathbf{f}(\mathbf{C}_t, \mathbf{p}(\mathbf{C}_t;\bm{\phi})) \Biggl) \\
    + \frac{\rho}{2} \Biggl|\!\Biggl| \mathbf{g}\Biggl(\frac{1}{T} \sum_{t=0}^{T-1} \mathbf{f}(\mathbf{C}_t, \mathbf{p}(\mathbf{C}_t;\bm{\phi})) \Biggl) \Biggr|\!\Biggr|^2
\end{multline}
In order to train the model parameter $\bm{\phi}_t$, we present an iteration duration $T_0$, which is equal to the number of time steps between consecutive model parameter updates. With a slight abuse of notation for time $t$, we define an iteration index $m \in \{0,1, 2, ..., M-1\}$, where $M = \lfloor T/T_0 \rfloor$ and model parameter is updated as follows.
\begin{equation} \label{phim_update_par}
    \bm{\phi}_m = \arg \max_{\bm{\phi} \in \bm{\Phi}} \mathcal{L}(\bm{\phi}, \bm{\mu}_m).
\end{equation}
The dual variables, $\bm{\mu}$ are updated recursively as 
\begin{multline} \label{mum_update_par}
    \bm{\mu}_{m+1} =  \Bigg[ \bm{\mu}_m - \rho ~ \mathbf{g}\Biggl(\frac{1}{T} \sum_{t=mT_0}^{(m+1)T_0-1} \mathbf{f}(\mathbf{C}_t, \mathbf{p}(\mathbf{C}_t;\bm{\phi})) \Biggl) \Bigg]^+
\end{multline}
 where the penalty parameter $\rho$, represents the learning rate for the dual variable update. The above routing algorithm can result in a feasible and near-optimal decision when run for a large enough number of time instants. 


 It can be noted that using regular dual descent algorithms like Primal-Dual methods may not guarantee a feasible set of routing decisions since the algorithm in \eqref{phim_update_par}, \eqref{mum_update_par} adapts the routing policy to the dual variables in each iteration. Given their efficient results, such parameterized policies have significant drawbacks which requires us to improve the approach to obtaining routing policies.

\section{The State Augmentation Algorithm} \label{sec:state_aug}
As discussed previously, the iterative routing optimization in \eqref{phim_update_par}, \eqref{mum_update_par} is feasible, they have certain shortcomings which make it unsuitable for practical applications. Firstly, the maximization of the Lagrangian duals require the future or non-causal knowledge of the network state i.e. it is not possible to get knowledge of the system at $t=kT_0$. It may be possible during the training phase but infeasible in the testing phase. The other challenge is convergence to near-optimal network performance. This is feasible when the time $T$ tends to infinity, thus the training iterations can not be stopped after a finite time step. In other words, there may not exist an iteration index $m$ for which $\bm{\phi}_m$ is optimal or feasible. Finally, the optimal set of model parameters in \eqref{phim_update_par} are required to be found at each time step for a different vector of dual variables $\bm{\mu}_m$, thereby increasing the computational complexity, more precisely in the execution phase.

The above challenges mandate the need to design an algorithm that does not require to retrain the model parameters $\bm{\phi}_m$ for any given set of dual variables $\bm{\mu}_m$. Hence we propose a state-augmented routing algorithm as in \cite{calvo2021state, naderializadeh2022state}, where we augment the network state $\mathbf{C}_t$ at each time step $t$ with the corresponding set of dual variables $\bm{\mu}_{\lfloor t/T_0 \rfloor}$. These dual variables are fed as simultaneous inputs along with the input node features to the GNN model, $A_i^k(t)$ to obtain the optimal routing policy. In other words, the GNN model in this case takes two input features, i.e. $F_0=2$. We present a different parameterization for the state-augmented routing policy, where the routing decisions $\mathbf{p}(\mathbf{C}_t)$ are represented in the form $\mathbf{p}(\mathbf{C}_t; \bm{\theta})$, where $\bm{\theta} \in \bm{\Theta}$ denotes the set of GNN filters tensors parameterized by the state-augmented routing policy. We start by defining the augmented Lagrangian of \eqref{eq:gnn_par} for a set of dual variables $\bm{\mu} \in \mathbb{R}_+^c$ as expressed in  \eqref{aug_lag1}. 
\begin{multline} \label{eq:aug_lag2}
    \mathcal{L}_{\bm{\mu}}(\bm{\theta}) = \mathcal{U} \Biggl(\frac{1}{T} \sum_{t=0}^{T-1} \mathbf{f}(\mathbf{C}_t, \mathbf{p}(\mathbf{C}_t;\bm{\theta}))\Biggl) \\
    + \bm{\mu}^T  \mathbf{g}\Biggl(\frac{1}{T} \sum_{t=0}^{T-1} \mathbf{f}(\mathbf{C}_t, \mathbf{p}(\mathbf{C}_t;\bm{\theta})) \Biggl) \\
    + \frac{\rho}{2} \Biggl|\!\Biggl| \mathbf{g}\Biggl(\frac{1}{T} \sum_{t=0}^{T-1} \mathbf{f}(\mathbf{C}_t, \mathbf{p}(\mathbf{C}_t;\bm{\theta})) \Biggl) \Biggr|\!\Biggr|^2
\end{multline}
If we try to plug in the functional values of the optimization problem in \eqref{eq:prob_state}, we obtain the augmented Lagrangian as follows.
\begin{multline} \label{eq:lagrangian_applied}
    \mathcal{L}_{\bm{\mu}} (\bm{\theta}) = \sum_k \sum_i \log \Biggl(\frac{1}{T} \sum_{t=0}^{T-1} a_i^k(t) \Biggl) \\
    + \bm{\mu}_i^k \Bigg[ \frac{1}{T} \sum_{t=0}^{T-1} \bigg( \sum_{j \in \mathcal{N}_i} r_{ij}^k(t) - \sum_{j \in \mathcal{N}_i} r_{ji}^k(t) - a_i^k(t) - z_i^k(t) \bigg)\Bigg] \\
    - \frac{\rho}{2} \Biggl|\!\Biggl| \sum_{j \in \mathcal{N}_i} r_{ij}^k(t) - \sum_{j \in \mathcal{N}_i} r_{ji}^k(t) - a_i^k(t) - z_i^k(t) \Biggr|\!\Biggr|^2
\end{multline}
where the Lagrangian multipliers $\bm{\mu}_i^k$ are associated with the constraint (\ref{eq:7b}). We keep the constraints \eqref{eq:7c} and \eqref{eq:7d} implicit. Next we consider the dual variables to be drawn from a probability distribution $p_{\bm{\mu}}$. This leads to defining the state-augmented routing policy as the one which maximizes the expectation of the augmented Lagrangian over the probability distribution of all parameters, \textit{i.e.}
\begin{equation} \label{eq:theta*_sa}
    \bm{\theta}^* = \arg \max_{\bm{\theta} \in \bm{\Theta}} \mathbb{E}_{\bm{\mu} \sim p_{\bm{\mu}}} \big[\mathcal{L}_{\bm{\mu}} (\bm{\theta}) \big]
\end{equation}
The above state-augmented policy parameterized by $\bm{\theta}^*$ enables us to determine the Lagrangian maximized routing decision $\mathbf{p}(\mathbf{C};\bm{\theta})$ for every dual variable iteration $\bm{\mu} = \bm{\mu}_m$. Using the above concept, the dual variable update for the iteration $m$ in \eqref{eq:theta*_sa} can be updated as below.
\begin{equation} \label{eq:mu_sa1}
    \bm{\mu}_{m+1} = \Bigg[ \bm{\mu}_m - \eta_{\bm{\mu}} \mathbf{g}\Biggl(\frac{1}{T_0} \sum_{t=mT_0}^{(m+1)T_0-1} \mathbf{f}(\mathbf{C}_t, \mathbf{p}^{\bm{\theta}}(\mathbf{C}_t;\bm{\theta}^*)) \Biggl) \Bigg]^+
\end{equation}
Plugging in the required function values for our optimization problem in \eqref{eq:prob_state}, the dual variable update for the execution phase can be given as,
\begin{multline} \label{eq:mu_sa2}
    \bigg[\bm{\mu}_i^k(t)\bigg]_{m+1} = \bigg[\bm{\mu}_i^k(t)\bigg]_m \\
    - \eta_{\bm{\mu}} \Bigg[ \frac{1}{T} \sum_{t=0}^{T-1} \Bigg( \sum_{j \in \mathcal{N}_i} r_{ij}^k(t) - \sum_{j \in \mathcal{N}_i} r_{ji}^k(t) - a_i^k(t) \bigg) \Bigg]
\end{multline}
The above equation helps us in learning a parameterized model in parallel while solving for the optimal state-augmented routing policy in \eqref{eq:theta*_sa}, eventually executing the gradient descent techniques of a standardized dual descent algorithm. It can also be induced from Theorem 2 in \cite{naderializadeh2022state} that the routing decisions made by the state-augmented algorithm in \eqref{eq:mu_sa1}, \eqref{eq:theta*_sa} are both feasible, and near-optimal. Since we have routing matrices resulting from the local information operated on channel capacity matrices, they can be best attributed to the operation on signals on graphs. 

\begin{subsection}{Implementation Scenario and Practical Assumptions}
The maximization in \eqref{eq:theta*_sa} is supposed to be carried out during the offline training phase. We resort to the gradient ascent technique to learn an optimal set of parameters $\bm{\theta}^*$ which can be saved after the completion of the training step and utilized later during the execution phase. During the training phase, we consider a batch of dual variables $\{ \bm{\mu}_b \}_{b=1}^{B}$ sampled randomly from the probability distribution $p_{\bm{\mu}}$. Thus the empirical form of the Lagrangian maximization in \eqref{eq:theta*_sa} can be reformulated as
\begin{equation} \label{eq:16}
    \bm{\theta}^* = \arg \max_{\bm{\theta} \in \bm{\Theta}} \frac{1}{B} \sum_{b=1}^{B-1} \mathcal{L}_{\bm{\mu}_b} (\bm{\theta})
\end{equation}
Just like a primal variable is maximized in primal-dual learning, the above maximization problem can be solved iteratively using the gradient ascent method. In order to initiate the above process, the model parameters were randomly initialized as $\theta_0$ and they can be updated over an iteration index $n=0,...,N_{train}-1$. The parameter update equation for $\bm{\theta}$ can be expressed as below,
\begin{equation} \label{eq:par_update_sa}
    \bm{\theta}_{n+1} = \bm{\theta}_{n} + \frac{\eta_\theta}{B} \sum_{b=1}^{B-1} \nabla_\theta \mathcal{L}_{\bm{\mu}_b} (\bm{\theta}_n)
\end{equation}
where $\eta_\theta$ delineates the learning rate for the model parameters $\theta$ or it can be denoted as the coefficients of the graph filter for the Graph Neural Network (GNN) parameterization discussed previously. Note that the model parameters here refer to the filter coefficients of the graph filters used in the GNN network. Subsequently the update of the primal variables $\bm{a,r}$ is updated by the GNN once the model parameters are updated using \eqref{eq:par_update_sa}. Once the model is trained, the final set of converged parameters are stored as $(\bm{\theta}^*)$. The generalization capability of the trained model can be improved for each set of dual variables, $\{\bm{\mu}_b\}_{b=0}^{B-1}$ in the batch, by randomly sampling a separate realization of the sequence of network states $\{\mathbf{C}_{b,t} \}_{t=0}^{T-1}$. This facilitates the model to be optimized over a family of network realizations and we summarize the above training procedure in  Algorithm \ref{alg:alg1}. 

\begin{algorithm*}
\caption{Training Phase for State-Augmented routing algorithm.}
\label{alg:alg1}
\begin{algorithmic}[1]
 \renewcommand{\algorithmicrequire}{\textbf{Input:}}
 \REQUIRE Number of training iterations $N_{train}$, batch size B, number of time steps, T, primal learning rate $\eta_\theta$
\STATE \textit{Initialisation} : $\bm{\theta}_0, q_i^k(0)$ 
\FOR{$n=0,...,N_{train} -1$}
\FOR{$b=0,...,B-1$}
\STATE Randomly sample$\bm{\mu}_b \sim p_{\mu}$
\STATE Randomly generate a sequence of network states $\{\mathbf{C}_{b,t} \}_{t=0}^{T-1}$
\FOR{$t=0,...,T-1$}
\STATE Generate routing decisions $\mathbf{p}(\mathbf{C}_{b,t},\mathbf{x}_b;\bm{\theta}_n))$
\STATE Obtain queue length using \eqref{queueupdate}
\ENDFOR
\STATE Calculate the augmented Lagrangian dual according to \eqref{eq:aug_lag2}, i.e.
\begin{multline*}
    \mathcal{L}_{\bm{\mu}_b}(\bm{\theta}) = \mathcal{U} \Biggl(\frac{1}{T} \sum_{t=0}^{T-1} \mathbf{f} \Bigl(\mathbf{C}_{b,t}, \mathbf{p}(\mathbf{C}_{b,t}, \mathbf{x}_b;\bm{\theta}_n) \Bigr)\Biggl) 
    + \bm{\mu}^T  \mathbf{g}\Biggl(\frac{1}{T} \sum_{t=0}^{T-1} \mathbf{f} \Bigl(\mathbf{C}_{b,t}, \mathbf{p}(\mathbf{C}_{b,t}, \mathbf{x}_b;\bm{\theta}_n) \Bigr) \Biggl) \\
    + \frac{\rho}{2} \Biggl|\!\Biggl| \mathbf{g}\Biggl(\frac{1}{T} \sum_{t=0}^{T-1} \mathbf{f} \Bigl(\mathbf{C}_{b,t}, \mathbf{p}(\mathbf{C}_{b,t}, \mathbf{x}_b;\bm{\theta}_n) \Bigr) \Biggl) \Biggr|\!\Biggr|^2
\end{multline*}
\ENDFOR
\STATE Update the model parameters according to (\ref{eq:par_update_sa})
\begin{equation*}
    \bm{\theta}_{n+1} = \bm{\theta}_{n} + \frac{\eta_\theta}{B} \sum_{b=1}^{B-1} \nabla_\theta \mathcal{L}_{\bm{\mu}_b} (\bm{\theta}_n)
\end{equation*}
\ENDFOR
\STATE $\bm{\theta}^* \leftarrow \bm{\theta}_{N_{train}}$
\renewcommand{\algorithmicrequire}{\textbf{Output:}}
 \REQUIRE Optimal model parameters $\theta^*$
\end{algorithmic}
\end{algorithm*}

The subsequent stage brings us to the execution phase where the dual variables are updated to engender the routing policies for the operating network. First the dual variables are initialized to zero. For any given time instant $\{t\}_{t=0}^{T-1}$, and a given network state $\mathbf{C}_t$, the routing decisions are generated using the state-augmented policies $\mathbf{p}(\mathbf{C}_t,\mathbf{x}_{\lfloor t/T_0 \rfloor};\bm{\theta}^*)$, which were obtained in Algorithm \ref{alg:alg1}. The dual variables are updated for every $T_0$ time step as per (\ref{eq:mu_sa2}). The key point of observation is the fashion in which the dual dynamics in (\ref{eq:mu_sa2}) make the satisfaction of the constraints tractable. As we described the execution phase in Algorithm 2, we realize the routing decisions at time instant $t$ help satisfy the constraints when the dual variables are minimized. On the contrary, increase in the value of dual variables imply that the constraints are not satisfied and the execution phase needs attention to tune some of the algorithm parameters. Both the algorithms follow similar steps to the ones in \cite{naderializadeh2022state} with some minor changes specific to our problem formulation.

\begin{algorithm*}
\caption{Execution Phase for State-Augmented routing algorithm.}
\label{alg:alg2}
\begin{algorithmic}[1]
 \renewcommand{\algorithmicrequire}{\textbf{Input:}}
 \REQUIRE Optimal model parameters $\theta^*$, sequence of network states $\{\mathbf{C}_{b,t} \}_{t=0}^{T-1}$, iteration time $T_0$, dual learning rate $\eta_\mu$
\STATE \textit{Initialisation} : $\bm{\mu}_0 \leftarrow 0, m \leftarrow 0$
\FOR{$t=0,...,T-1$}
\STATE Generate routing decisions $\mathbf{p}_t$ := $\mathbf{p}(\mathbf{C}_t,\mathbf{x}_m;\bm{\theta}^*))$
\STATE Obtain queue length using (\ref{queueupdate})
\IF{$(t+1)$ mod $T_0$ = 0}
\STATE Update the dual variables according to (\ref{eq:mu_sa1})
\begin{equation*} 
    \bm{\mu}_{m+1} = \Bigg[ \bm{\mu}_m 
    - \eta_{\bm{\mu}} \mathbf{g}\Biggl(\frac{1}{T_0} \sum_{t=mT_0}^{(m+1)T_0-1} \mathbf{f} \Bigl(\mathbf{C}_t, \mathbf{p}(\mathbf{C}_t,\mathbf{x}_m;\bm{\theta}^*) \Bigr) \Biggl) \Bigg]^+
\end{equation*}
$m \leftarrow m+1$
\ENDIF
\ENDFOR
\renewcommand{\algorithmicrequire}{\textbf{Output:}}
 \REQUIRE Sequence of routing decisions $\{\mathbf{p}_t\}_{t=0}^{T-1}$
\end{algorithmic}
\end{algorithm*}
\end{subsection}

\section{Experimental Observations \\ \& Numerical Results} \label{sec:results}
\subsection{Communication Network Architectures}
We initiate the experiments by generating random geometric network graphs of size $N=|\mathcal{V}|$. $k$-Nearest Neighbor ($k$-NN) method was followed to randomly generate $N$ communicating nodes in a unit circle. We considered the $k=4$ for all our simulations involving random graphs. 
For the architecture, we use a 3-layer GNN with $F_0 = 2, F_1=32$, and $F_2=8$ features. We utilize the ADAM optimizer with a primal learning rate of $\eta_{\bm{\theta}} = 0.005$ to optimize the primal model parameters, and we consider the penalty term $\rho = 0.005$, which is decayed exponentially to optimize the dual variables. We consider the following time steps for our simulations i.e. $T=100$ and $T_0=5$. The above conditions have been assumed for a time varying channel capacity whereas we assume a constant channel during our experiments, i.e. we need to find the policy $\mathbf{p}(\mathbf{C},\mathbf{x};\bm{\theta}^*)$ for $\mathbf{C} \in \mathcal{C}$. We generate a total of 128 training samples and 16 testing samples for each network size with a batch size of 16 samples. We run the training for 40 epochs and the dual variables for the training phase were drawn randomly from the $U(0,1)$ distribution. 
\subsection{Performance Across Different Unparameterized Algorithms}
\begin{figure}[htp]
    \centering   
    \includegraphics[width=3.5in]{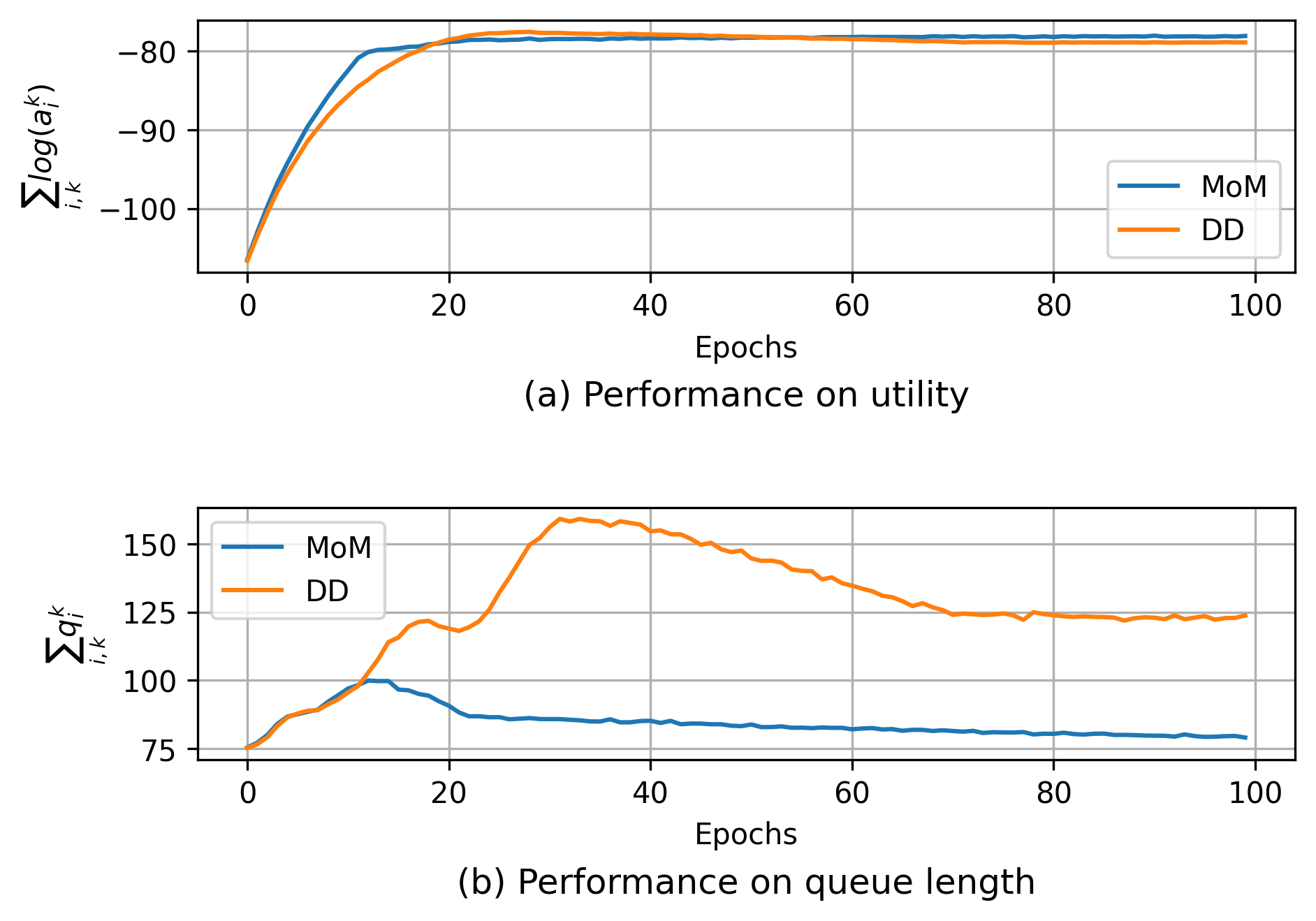}
    \caption{Comparison of two different unparameterized methods of MoM and Dual Descent (DD) for a network with 10 nodes and 5 flows, where the network is run only for a single time step.}
    \label{Fig:unpar_algo}
\end{figure}

Initially, we compare the performances of two unparameterized optimization methods by considering networks of size $N=10$ nodes and $K=5$ flows. We run the two methods for 100 epochs with $T=1$ in order to reduce the total number of optimization variables. Fig~\ref{Fig:unpar_algo} shows the performance for the 2 unparameterized methods of MoM, and Dual Descent (DD). It can be clearly observed from the utility plot that DD converges to the optimal solution slower while the augmented Lagrangian or MoM performs the best in terms of maximizing the utility and minimizing the queue lengths. The same can be attributed to queue length, as slow convergence amounts to piling up of more packets in the queues at the nodes.
\subsection{State-Augmentation vs ADMM}
\begin{figure}[h]
    \centering   
    \includegraphics[width=3.5in]{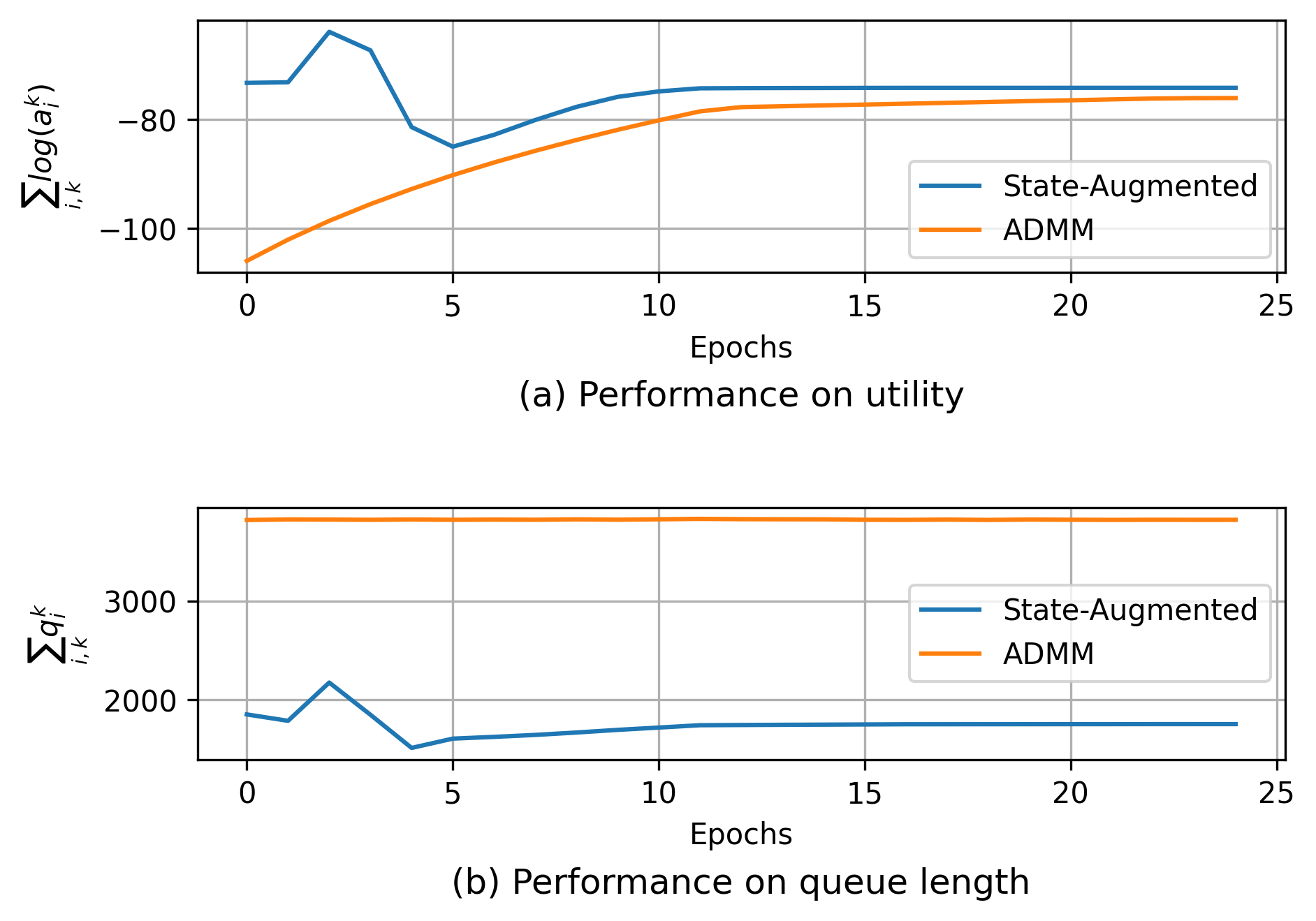}
    \caption{Performance comparison between unparameterized ADMM and the proposed parameterized state-augmented method using GNNs for networks with 10 nodes and 5 flows, run over $T=100$ time steps.}
    \label{Fig:par_vs_unpar}
\end{figure}
Next, we train the proposed state-augmented GNN-based model with random networks having $N=10$ nodes and $K=5$ flows, executed over $T=100$ time steps. We compare its performance with that of ADMM. As it can be seen from 
Fig~\ref{Fig:par_vs_unpar}, the proposed method performs better than the non-learning method of ADMM. The key point of observation is that while the proposed state-augmented model approaches the optimal utility attained by ADMM in Fig~\ref{Fig:par_vs_unpar}(a), it outperforms ADMM in terms of queue length in Fig. \ref{Fig:par_vs_unpar}(b).
\begin{figure}[t]
    \centering   
    \subfloat[\centering Performance on Utility \label{3(a)}]{%
    \includegraphics[width=3.5in]{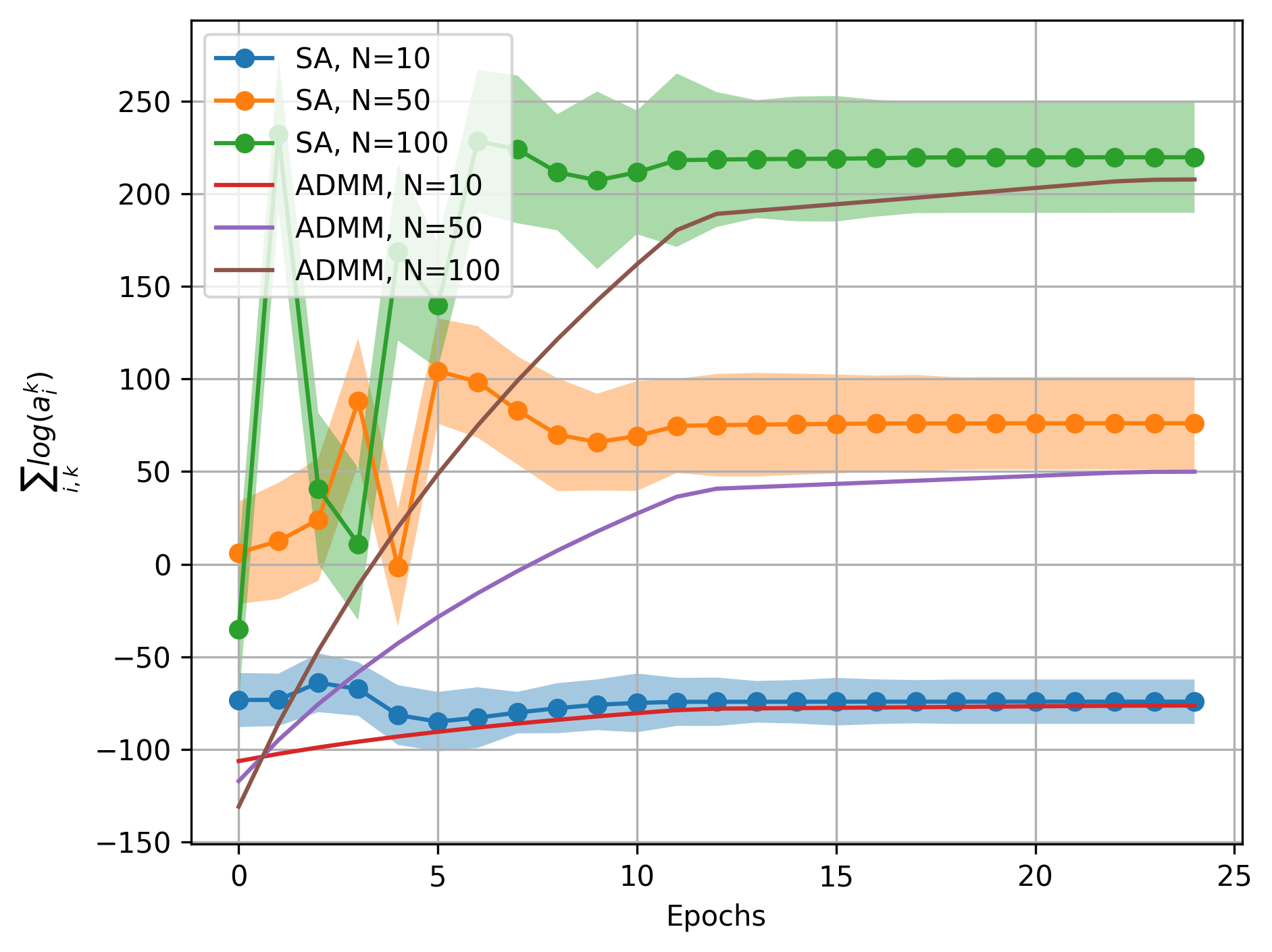}}
    \hfill
    \subfloat[\centering Performance on Queue length \label{3(b)}]{%
    \includegraphics[width=3.5in]{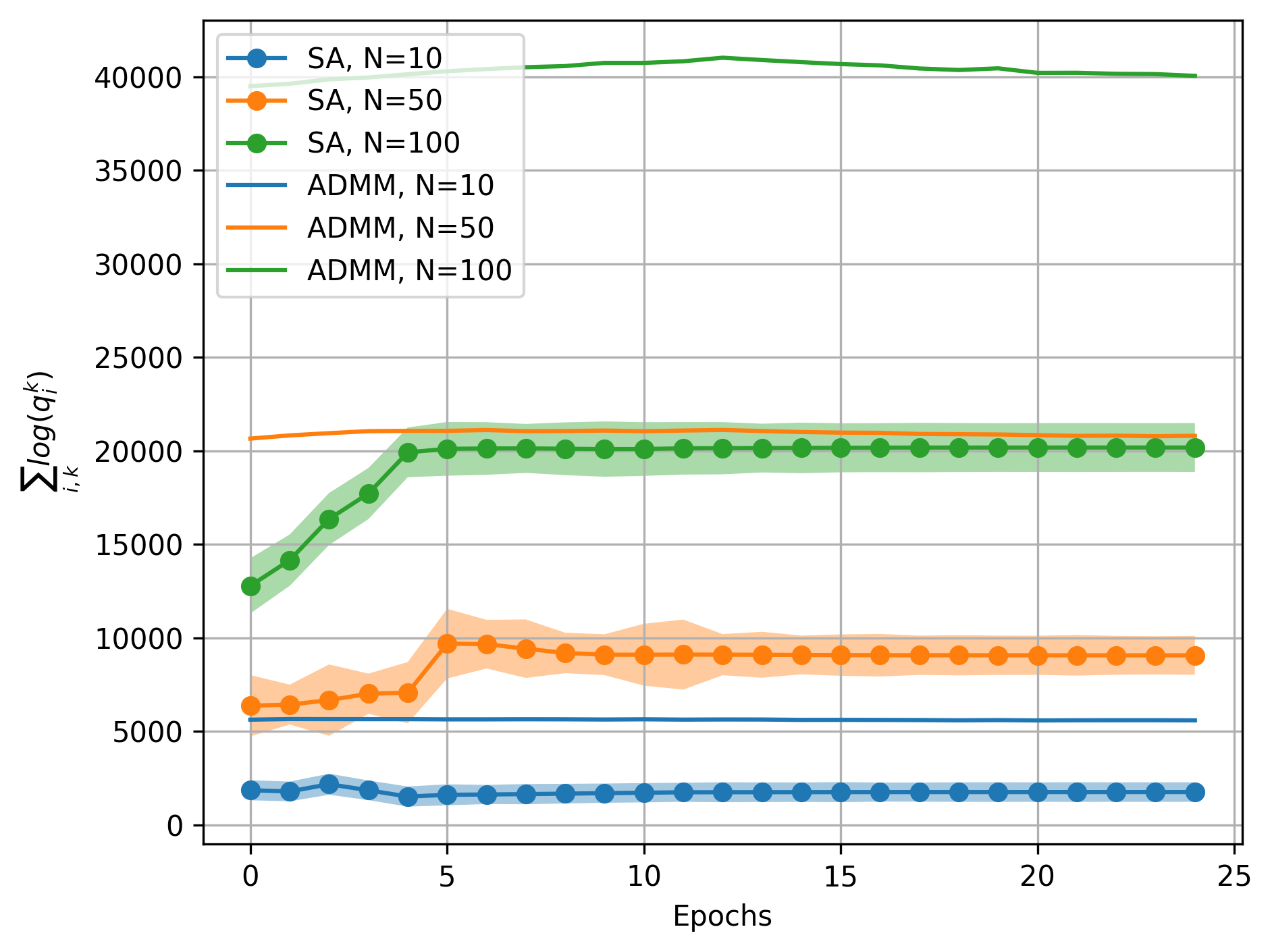}}
    \caption{Performance comparison of state augmentation and ADMM algorithms for networks with 5 flows and $N\in\{10, 50, 100\}$ nodes.}
    \label{Fig:sa_vs_admm_nodes}
\end{figure}

Now, we vary the number of nodes in the network and observe the behaviors of the two algorithms. As the number of nodes are increased from 10 to 100, we observe an increase in the utility performance of both State Augmentation (SA) and ADMM. Although the parameterized algorithms has more variability, the average of the GNN based algorithm still performs close to the non-learning ADMM algorithm (Fig.~\ref{Fig:sa_vs_admm_nodes}a). Similar statistics can be observed for the queue length stability in both the learning algorithms (Fig.~\ref{Fig:sa_vs_admm_nodes}b). 
\begin{figure*}
    \centering   
    \subfloat[\centering Performance on Utility \label{4(a)}]{%
    \includegraphics[width=3.5in]{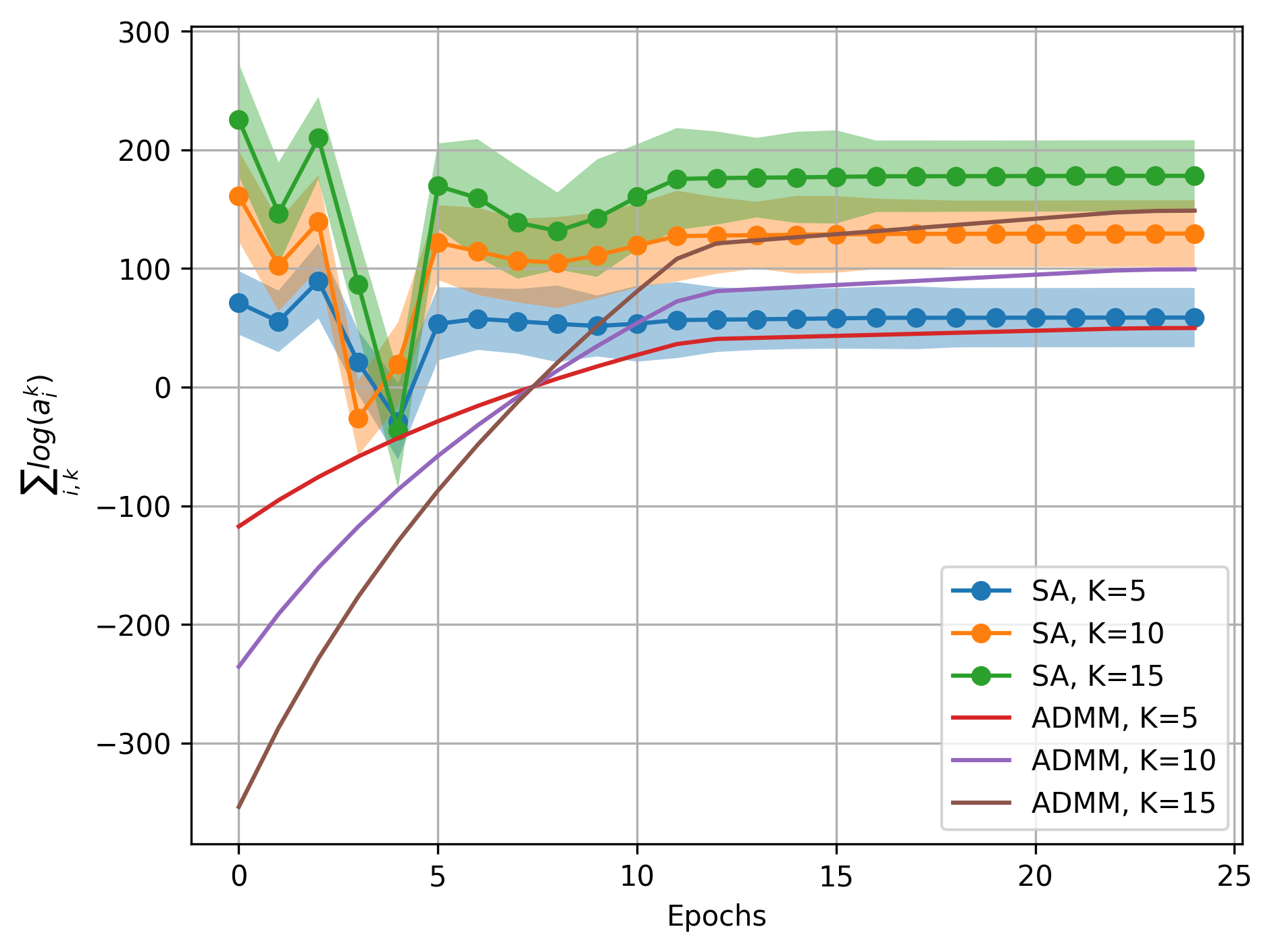}}
    \hfill
    \subfloat[\centering Performance on Queue length \label{4(b)}]{%
    \includegraphics[width=3.5in]{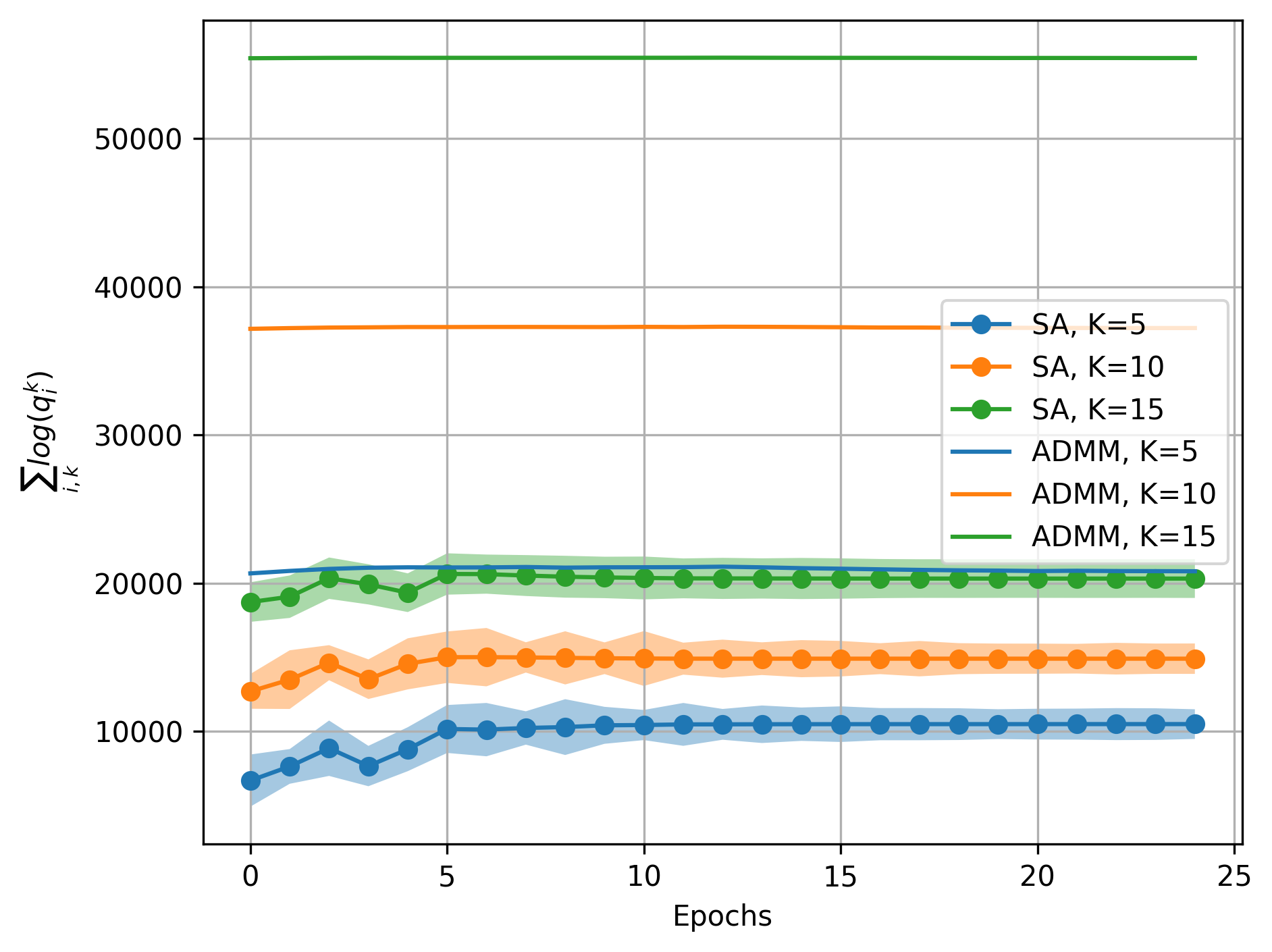}}
    \caption{Performance of state augmentation and ADMM algorithms for network with 50 nodes and $K\in\{5, 10, 15\}$ flows.}
    \label{Fig:sa_vs_admm_flows}
\end{figure*}
Furthermore, we test another scenario where the number of nodes in the network was kept fixed at 50 while the number of flows were increased from 5 through 15. We observe in Fig~\ref{Fig:sa_vs_admm_flows}(a) the performance of both algorithms increases with an increase in the number of flows while state augmentation still performs close to the optimal performance provided by ADMM. The queue size also increases with an increase in the number of flows as it facilitates more flows of packets in the network when the number of flows is increased. These results demonstrate the superior \emph{scalability} of the proposed GNN-based solutions vs. conventional optimization methods.

\begin{figure}
    \centering   
    \includegraphics[width=3.5in]{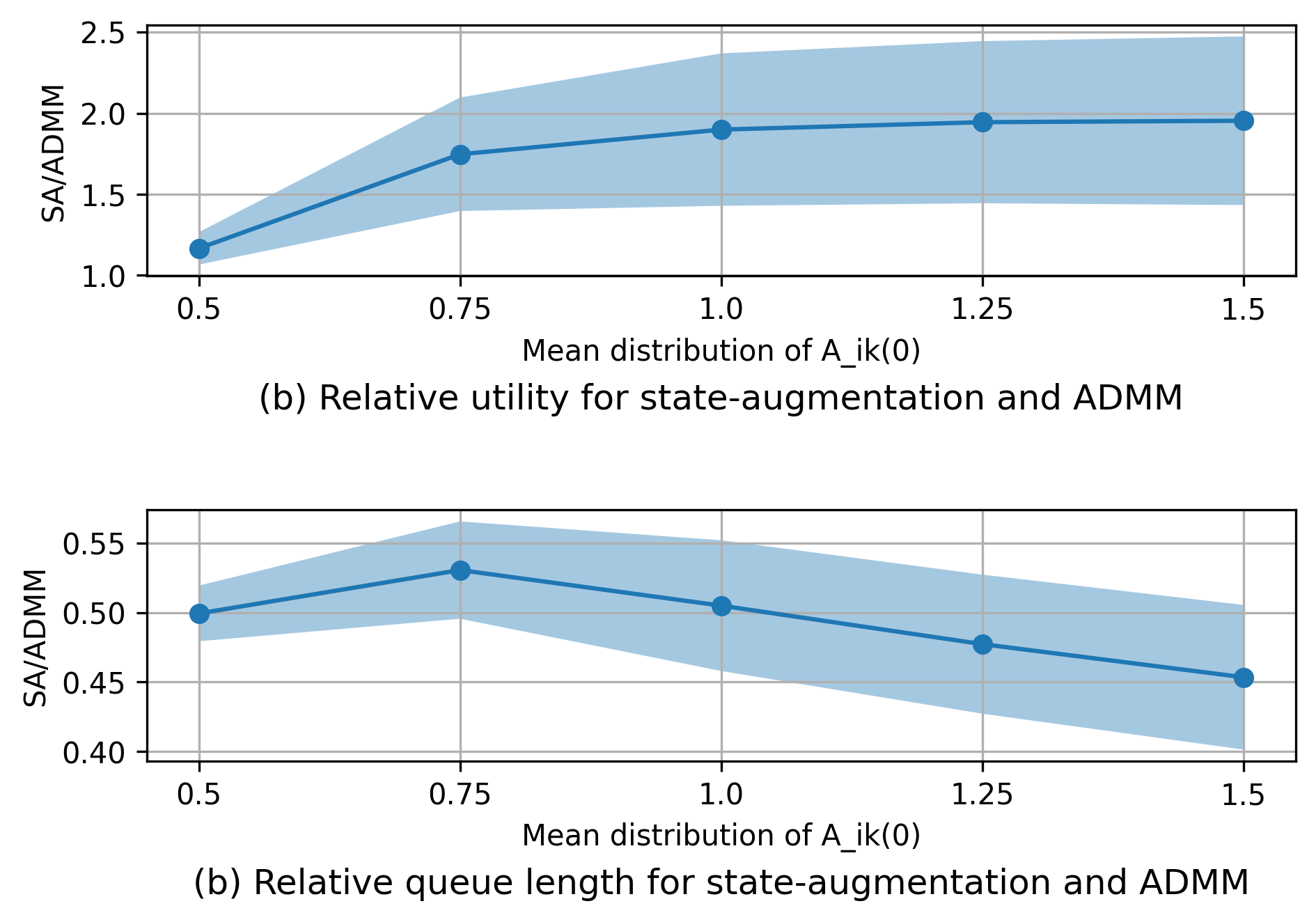}
    \caption{Relative performance of state-augmented algorithm with ADMM for a network with 50 nodes and 5 flows.}
    \label{Fig:rel_comp}
\end{figure}

The next point of consideration is the relative performance of state augmentation algorithm with the ADMM algorithm. We use the relative utility and queue length comparison for a network with 50 nodes and 5 flows. The mean value of the random input to the GNN, $A_i^k(t)$ is varied for 5 different traffic situations. As Fig.~\ref{Fig:rel_comp}(a) depicts, the relative performance of state augmentation based learning increases with an increase in the value of input traffic, $A_i^k(t)$. This implies higher input traffic is better processed by the GNN layers to pump more packets to their destination which inversely improves the queue length stability as evident from Fig.~\ref{Fig:rel_comp}(b). The variability in the shaded region is due to the state augmentation algorithm which makes use of the multiple GNN parameters. 
\subsection{Stability to Perturbation}

In order to test the stability property of GNNs, we provided perturbations to half of the nodes in a network which shifted their original location by 20\%. This leads to unanimous addition or reduction of edges in the network thereby creating a new graph for each sample. 
\begin{figure}[h]
    \centering   
    \includegraphics[width=3.5in]{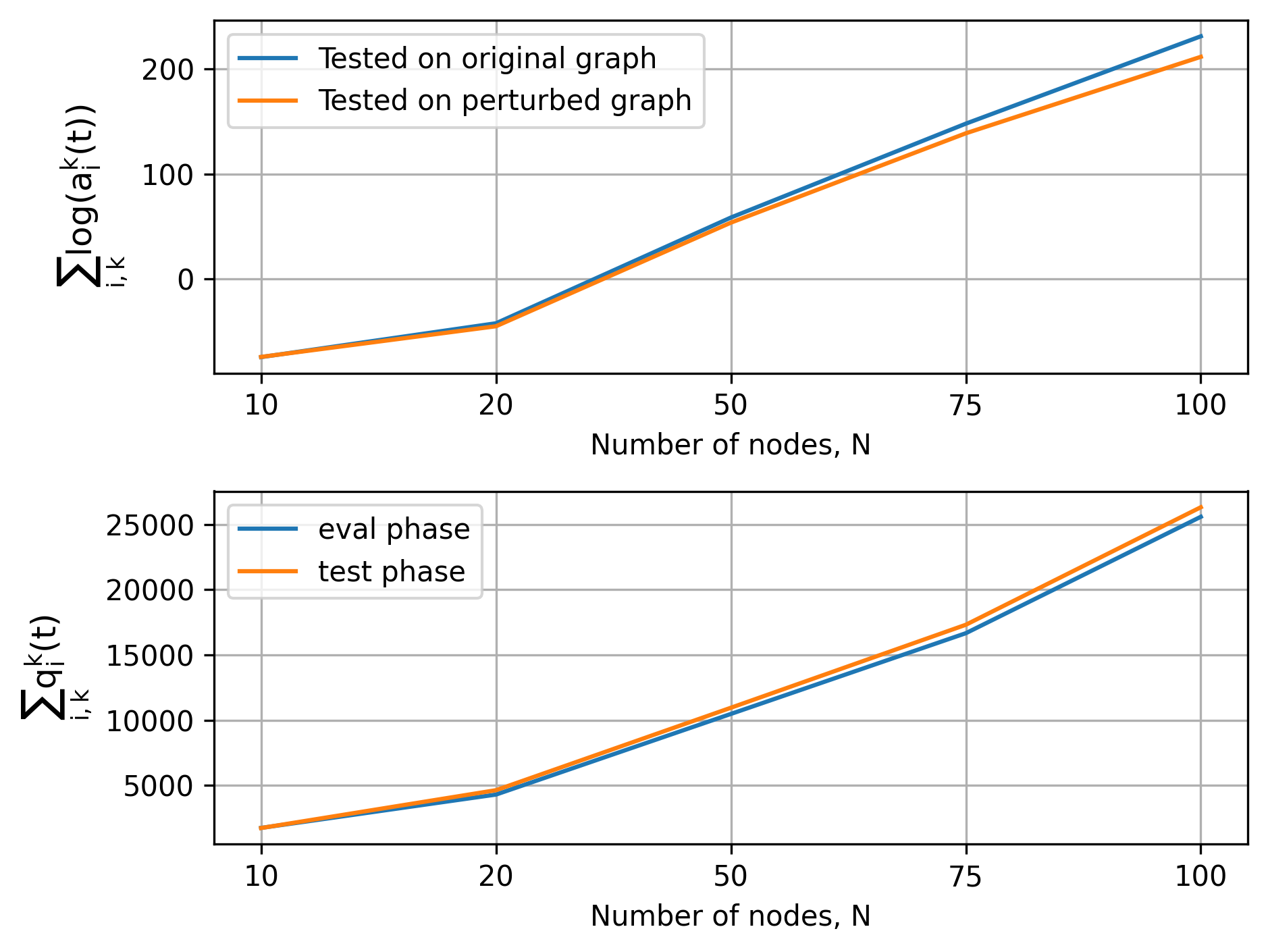}
    \caption{Performance of state augmentation based GNN to perturbation.}
    \label{Fig:perturb}
\end{figure}
As it can be seen in Fig.\ref{Fig:perturb}, when compared to the original graphs, the GNN model performs quite well on the perturbed graphs, hence supporting the stability property. It is to be noted that there can be minor deviations as the number of nodes increases because the number of changes in edge creation also varies substantially. We can consider an application of the above case in a varying network situation like flocking in multi agent systems where the nodes keep changing locations at every time step.

\subsection{Transferability to Unseen Graphs of various sizes}
As discussed in section~\ref{sec:gnn_par}, another advantage of GNNs is the size-invariance property where they can be seamlessly executed on network of different size which were previously not seen in the entire training phase. As a first approach we trained a state-augmented GNN model with $N=20$ nodes and $K=5$ flows. Then it was tested on graph with different sizes ranging from 10 nodes to 100 nodes.
\begin{figure}[h]
    \centering   
    \includegraphics[width=3.5in]{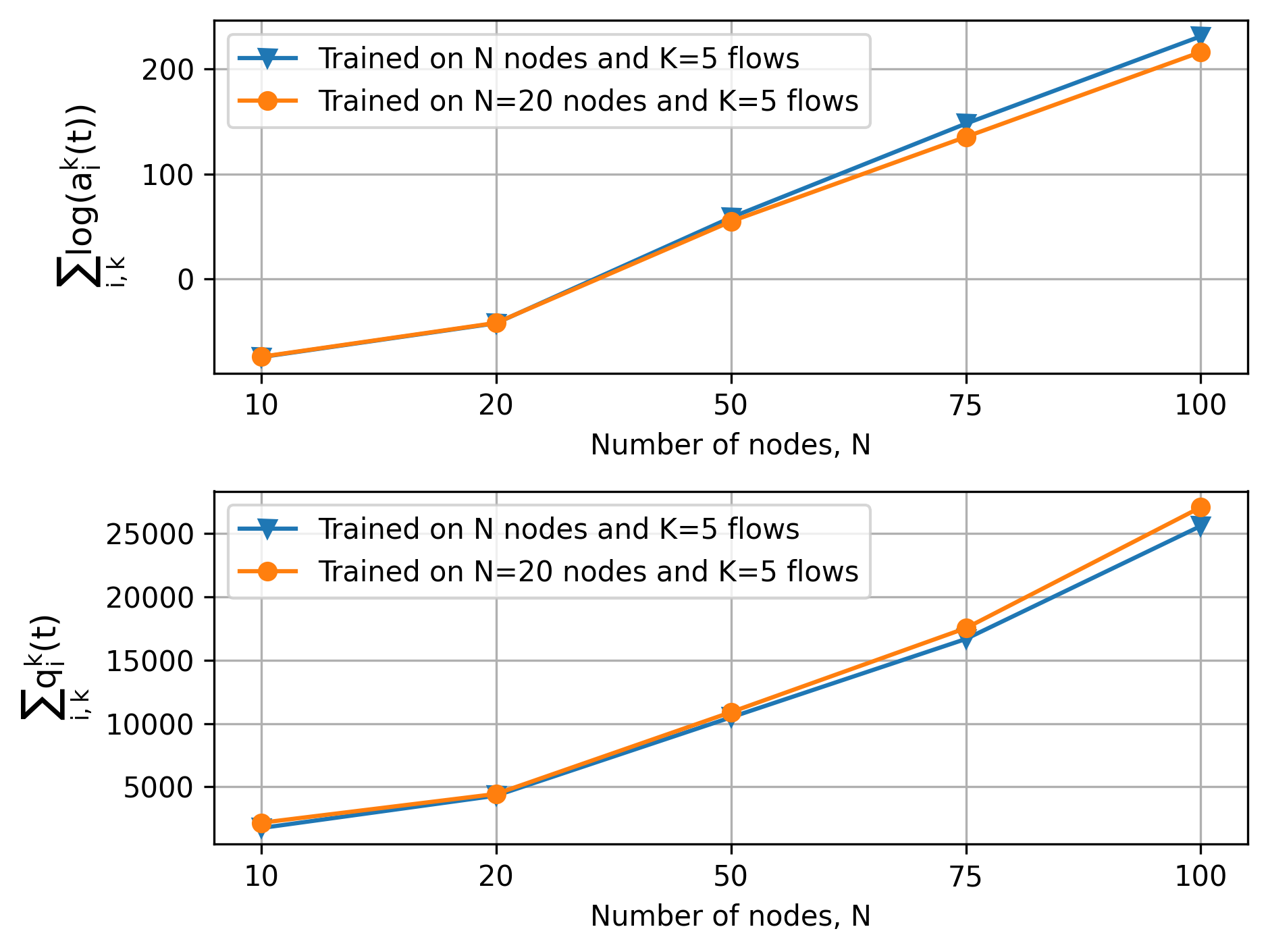}
    \caption{Transferability of the proposed state-augmented algorithm on networks with different nodes while they were trained on a network with 20 nodes and 5 flows.}
    \label{Fig:transfer_nodes}
\end{figure}
Fig.~\ref{Fig:transfer_nodes} shows the above transferability of a 50 node network under our proposed state-augmented algorithm to network of unseen sizes. The performance for transferability was found to be better for graphs of higher sizes while there was a minor increase in the queue length. The point for testing with 50 nodes also shows the stability of the GNN to other random graphs of the same size.

We perform similar experiments to observe the transferability by training a GNN model with $N=50$ nodes and $K=10$ flows. Now we test the same model with a network of $N=50$ nodes but different flows.
\begin{figure}[h]
    \centering   
    \includegraphics[width=3.5in]{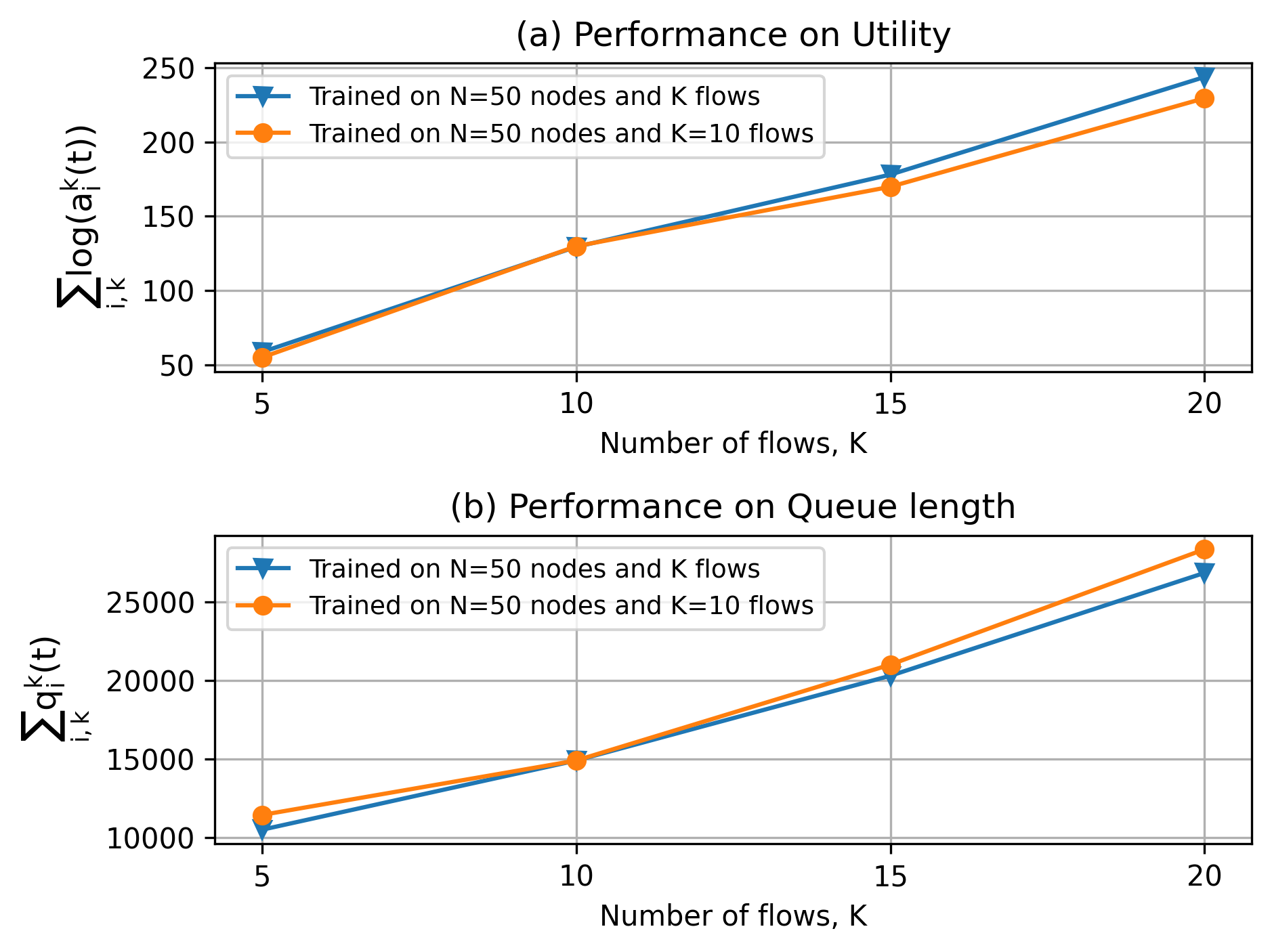}
    \caption{Transferability of the proposed state-augmented algorithm on networks with different flows while they were trained on a network with 50 nodes and 10 flows.}
    \label{Fig:transfer_flows}
\end{figure}
The transferability is satisfied again for the case of varying flows as depicted in Fig.\ref{Fig:transfer_flows}. The utility and the queue length stability perform close to the performance of the actual model trained on the respective graphs. Moreover the point for 10 flows shows the stability of the GNN to other random graphs of the same size and with same number of flows. This property of GNNs can be very helpful to train smaller networks offline and test them on larger networks online which will save considerable computational expenses.

\subsection{Dual variable and Queue length stability}
\begin{figure*}[htp]
\centering   
    \subfloat[\centering Behavior of dual variable \label{9(a)}]{%
    \includegraphics[width=3.5in]{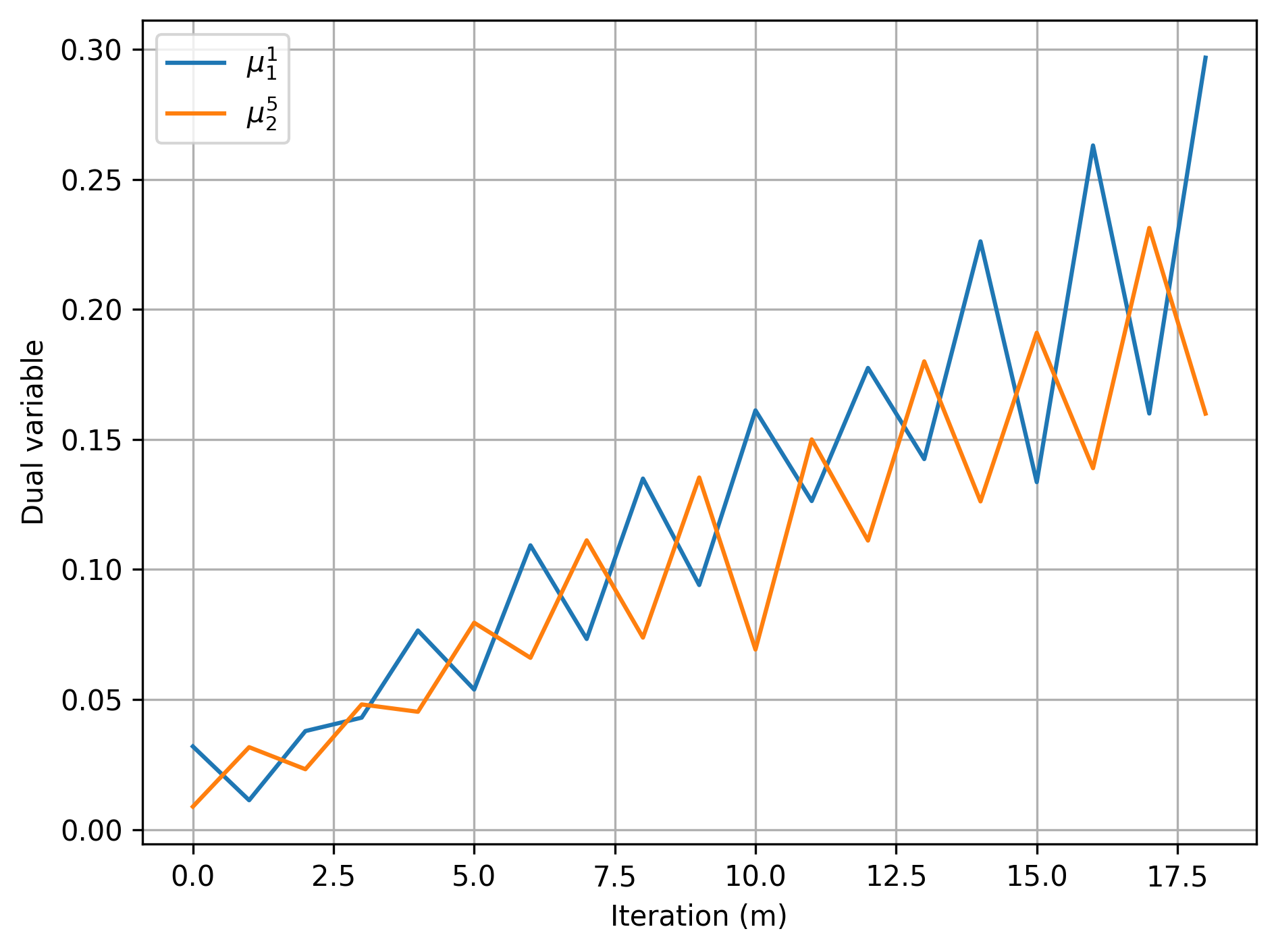}}
    \hfill
    \subfloat[\centering Queue length at a sample node \label{9(b)}]{%
    \includegraphics[width=3.5in]{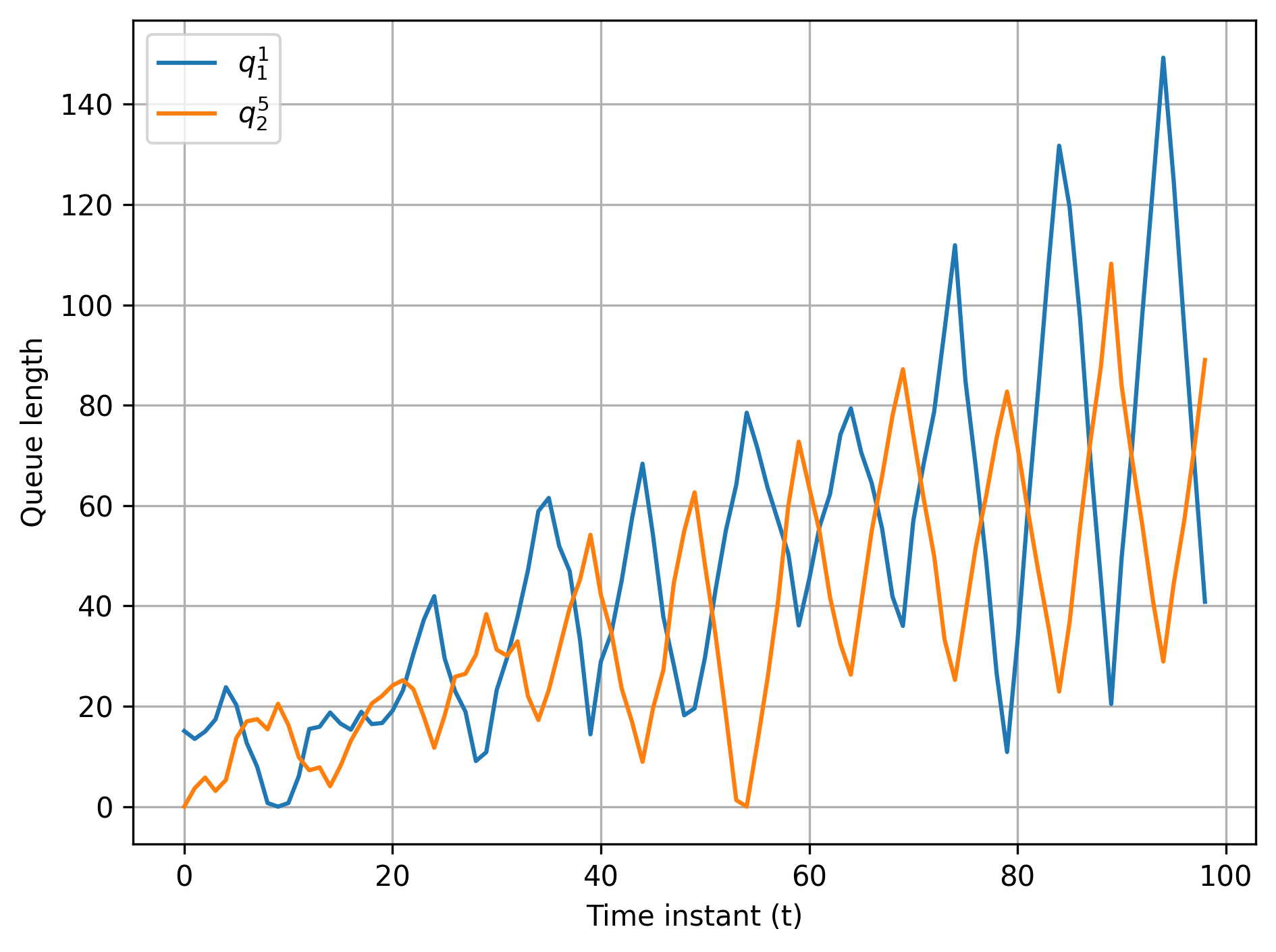}}
\caption{Behavior of a dual variable and queue length stability for an example node in a network with 50 nodes and 5 flows.}
\label{Fig:dual_perform}
\end{figure*}
For a general case of network samples with $N=10$, we train the model for $T=100$ and observe the behavior of dual variables and its relationship to the network performance. We plot the queue length at each time step $t$ and the corresponding dual variables $\bm{\mu}_{\lfloor t/T_0 \rfloor}$.  Fig.~\ref{Fig:dual_perform}(b) shows the queue length stability for two example nodes in the network over the course of $T=100$ time instants. It is observed that their corresponding queue lengths at the node increases as the dual variables increase with time. Once the dual variables converge to their optimal values, the queue length stabilizes to the best possible values.  

\begin{figure}[htp]
    \centering   
    \subfloat[\centering Nsfnet \label{10(a)}]{%
    \includegraphics[width=0.5\linewidth]{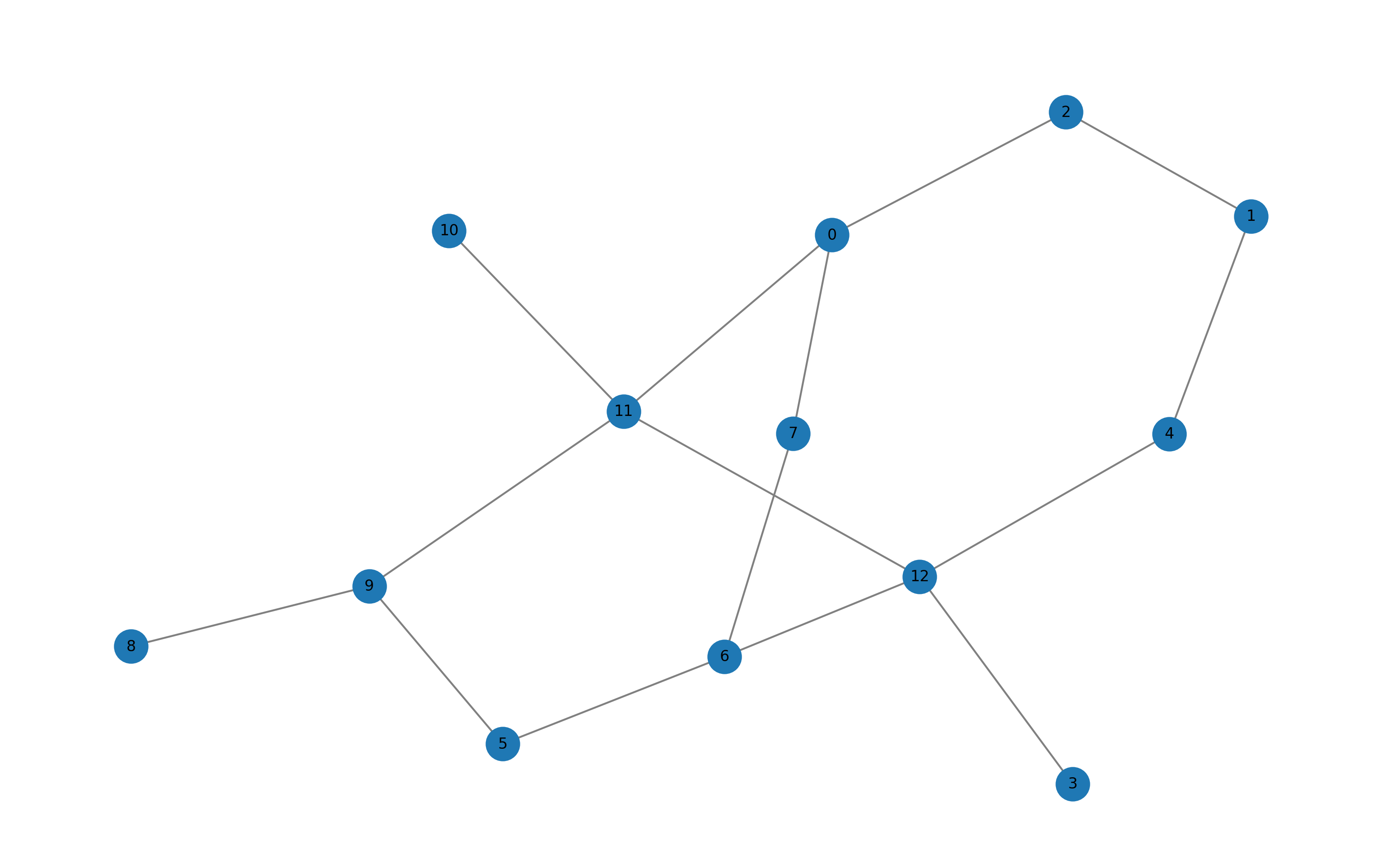}}
    \hfill
    \subfloat[\centering Missouri \label{10(b)}]{%
    \includegraphics[width=0.5\linewidth]{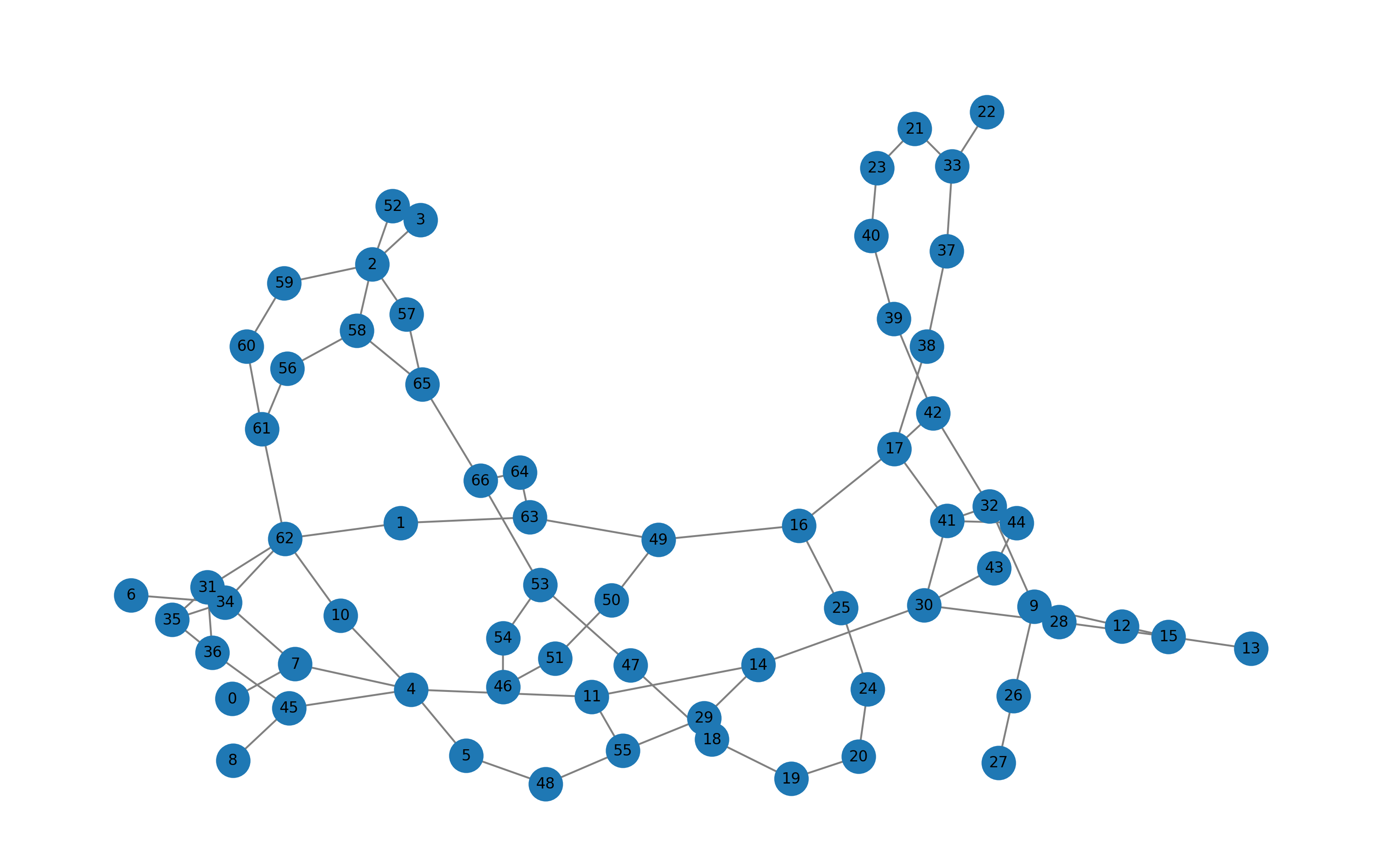}}
    \hfill
    \subfloat[\centering Sinet \label{10(c)}]{%
    \includegraphics[width=0.5\linewidth]{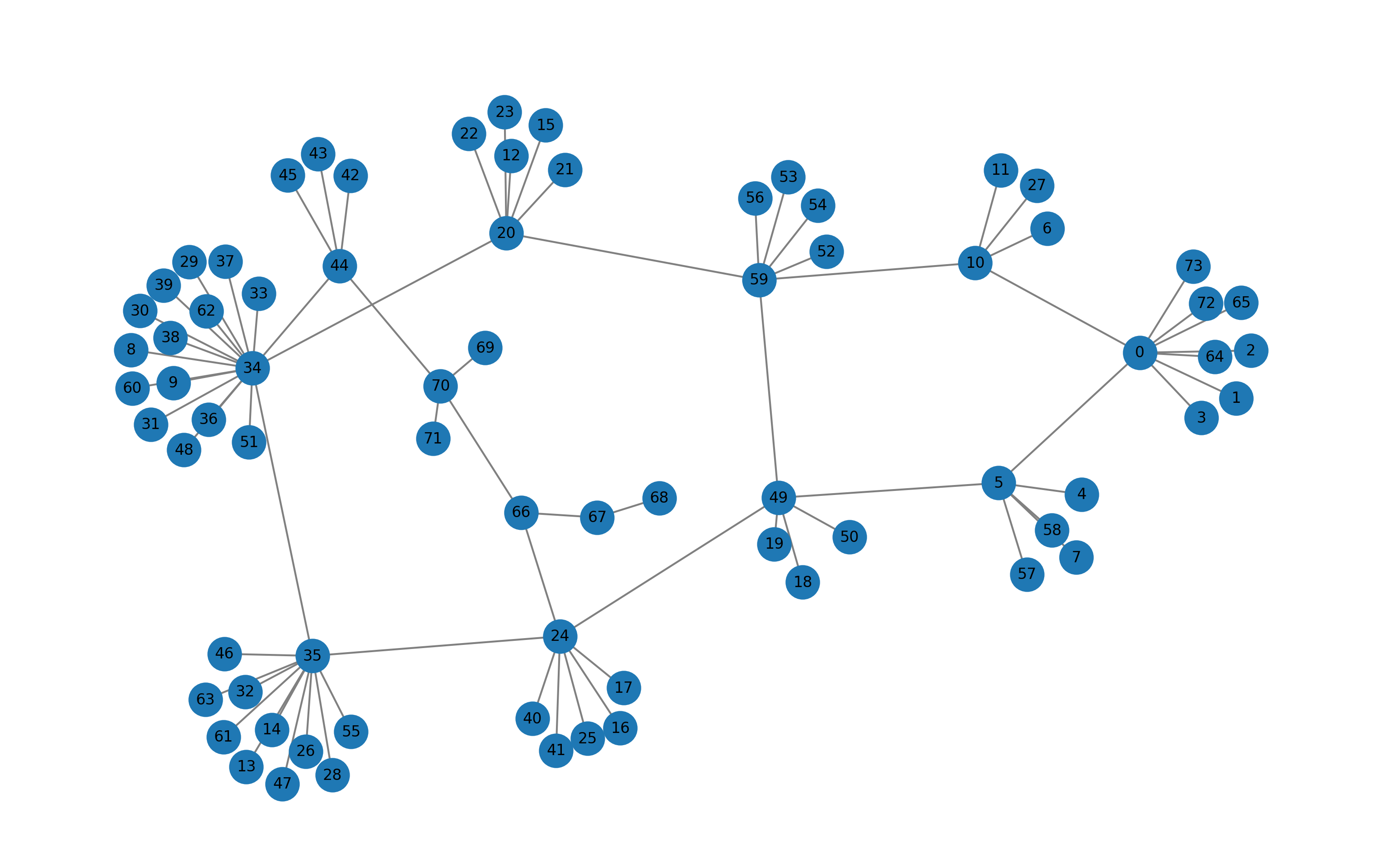}}
    \hfill
    \subfloat[\centering Interoute \label{10(d)}]{%
    \includegraphics[width=0.5\linewidth]{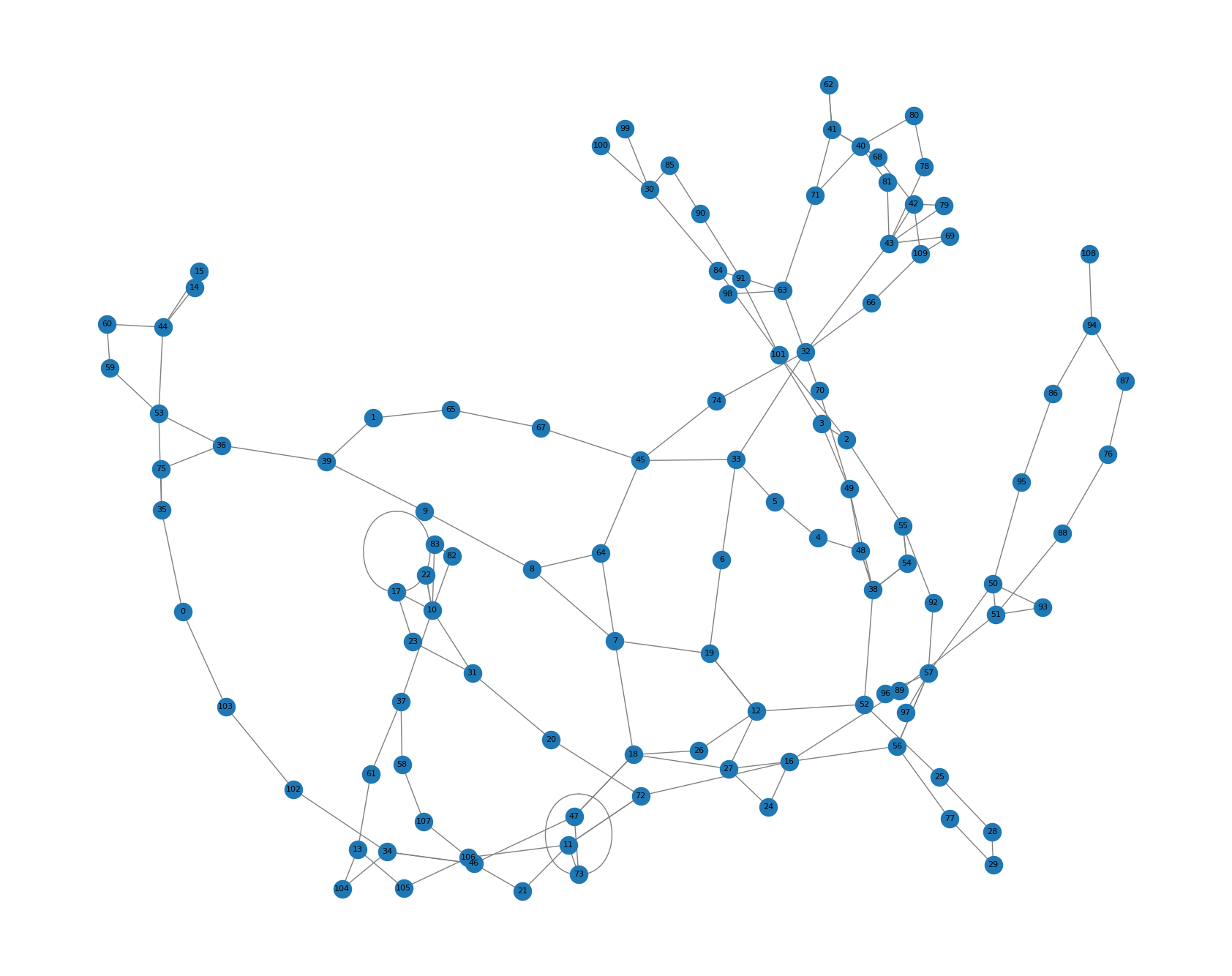}}
    \caption{Network topology graphs used from the Topology-Zoo dataset}
    \label{Fig:real_topology}
\end{figure}
\subsection{Performance on Real Time Network Topology Graphs}
\begin{figure*}
    \centering   
    \subfloat[\centering Performance on Utility \label{11(a)}]{%
    \includegraphics[width=3.5in]{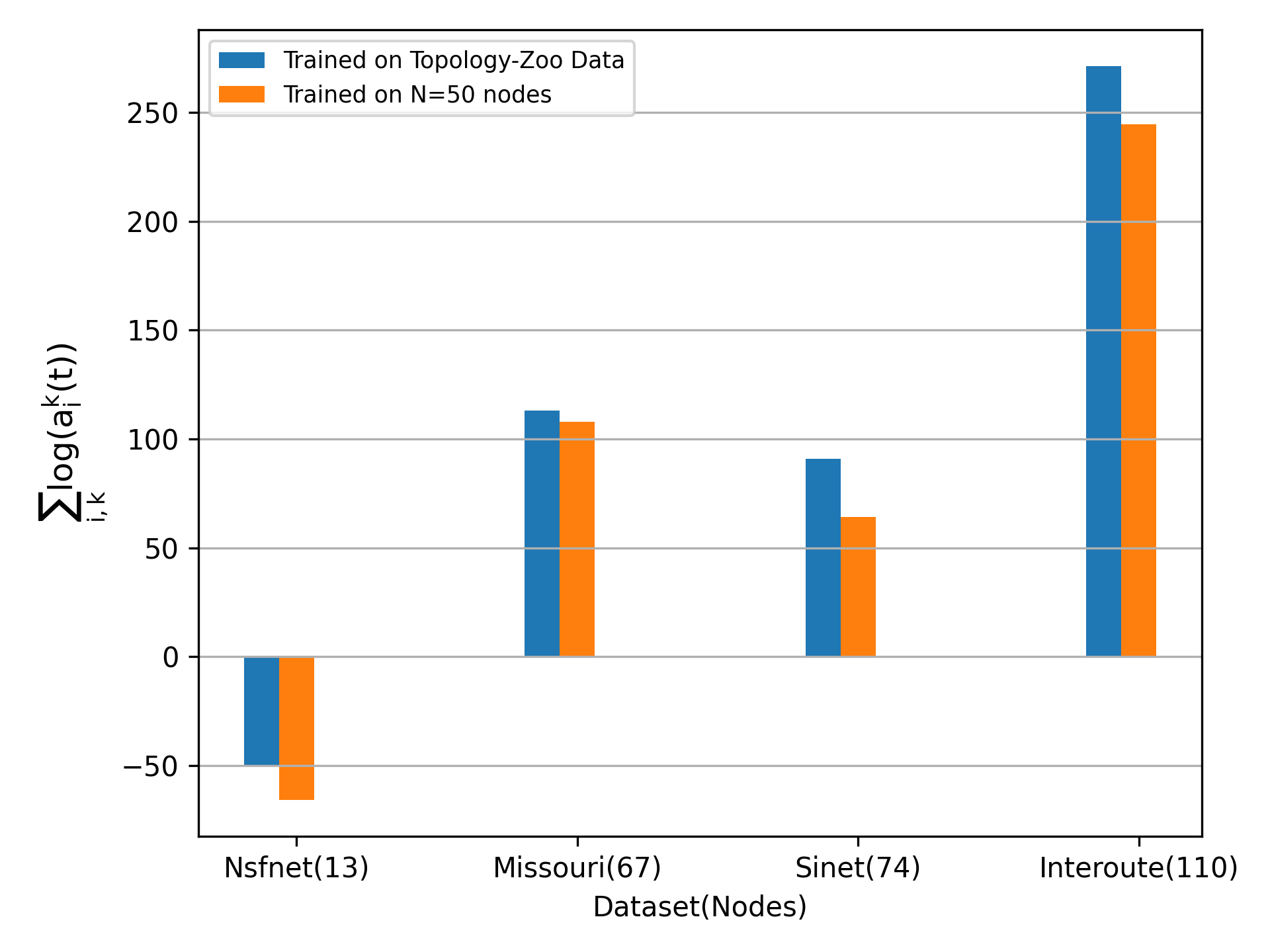}}
    \hfill
    \subfloat[\centering Performance on Queue length \label{11(b)}]{%
    \includegraphics[width=3.5in]{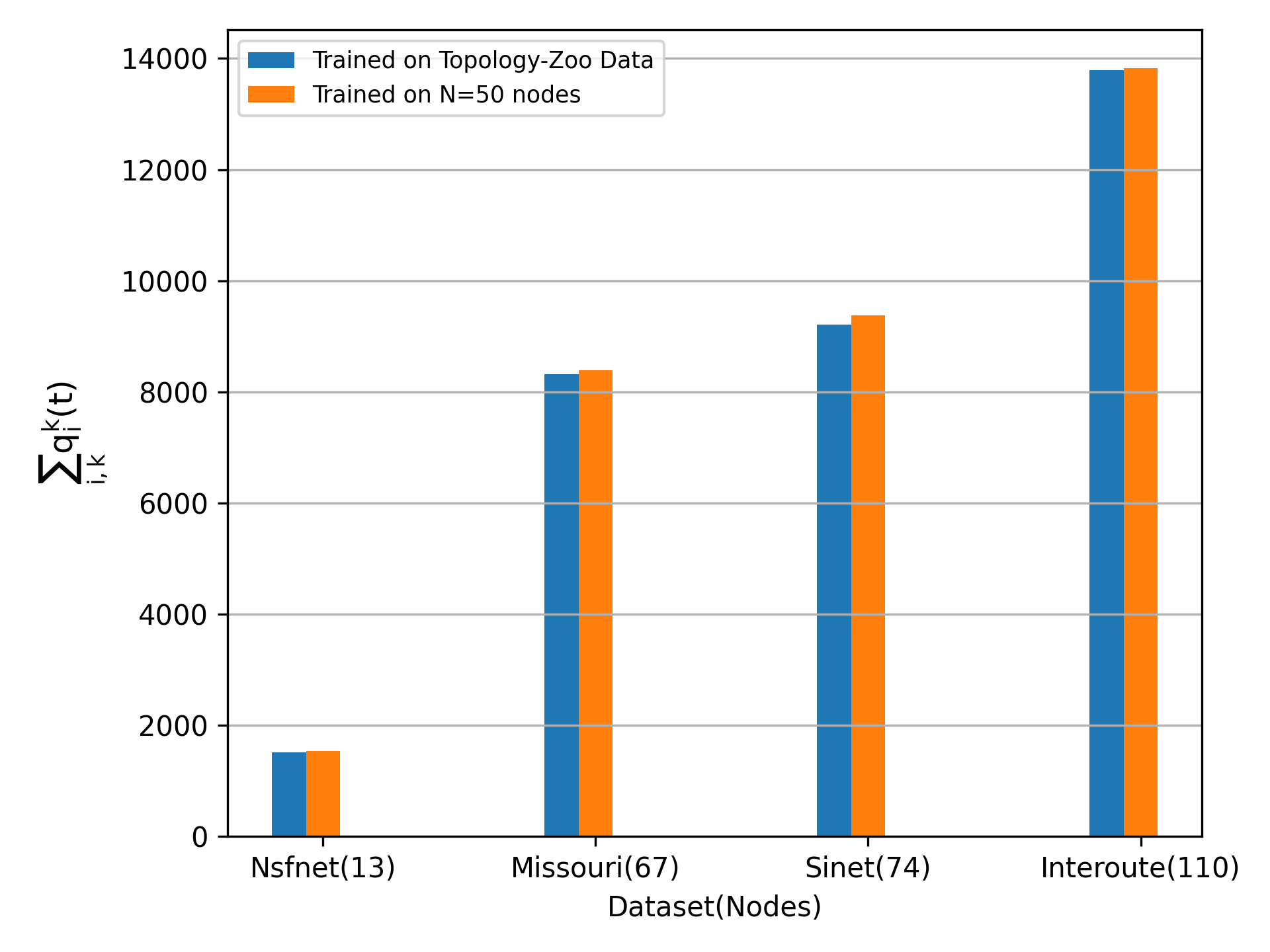}}
    \caption{Performance of State-Augmentation based routing policy on real world network topology}
    \label{Fig:real_data}
\end{figure*}

We finally evaluate the performance of the proposed method
on several real-world network topologies from the Internet Topology Zoo dataset \cite{knight2011internet}. We trained the models on 4 different network topologies as shown in the Fig.~\ref{Fig:real_topology} for 5 flows. Subsequently we transferred a model trained on 50 nodes and 5 flows using random data to test the robustness of the GNN architecture. Fig.~\ref{Fig:real_data} shows even if the trained models on the corresponding network topologies perform well, the transferred model which was trained on random data previously performs close to the former. However, if the network structure during testing deviate from those encountered during training, the model's transferability might be impacted. A particular case of performance can be seen in the Sinet graph where the transference does not perform well as compared to the others which is due to the cluster based structure of the network. Thus we can say that our proposed model is generalized to perform well on a wide variety of data sets and network topologies.

\section{Conclusion} \label{sec:conclusion}
In this paper, we considered the problem of routing information packets in a communication network where the objective was to maximize a network level utility function subject to various constraints. We carried out experiments via both learning and non-learning based approaches. The results in learning paradigm showed that they outperform non-learning methods although there was a cost involved in training the models initially. Considering the case of unparameterized learning, it was found to provide feasible and near optimal solutions but they have multiple challenges which include running them for an infinite number of time steps and the model parameters are required to be optimized for any set of dual variables at every time instant. Hence the State-Augmentation based routing algorithm was proposed to alleviate the above challenges of unparameterized learning. The experimental procedure and observation plots proved that the state augmentation based learning leads to feasible and near-optimal sequences of routing decisions, for near-universal parameterizations. Finally we used Graph Neural Network (GNN) architectures to parameterize the State-Augmentation based routing optimization which showcased the stellar properties of stability and transference to other networks including topologies from the real world datasets.

\bibliographystyle{IEEEtran}
\bibliography{references}

\begin{thebibliography}{10}
\providecommand{\url}[1]{#1}
\csname url@samestyle\endcsname
\providecommand{\newblock}{\relax}
\providecommand{\bibinfo}[2]{#2}
\providecommand{\BIBentrySTDinterwordspacing}{\spaceskip=0pt\relax}
\providecommand{\BIBentryALTinterwordstretchfactor}{4}
\providecommand{\BIBentryALTinterwordspacing}{\spaceskip=\fontdimen2\font plus
\BIBentryALTinterwordstretchfactor\fontdimen3\font minus
  \fontdimen4\font\relax}
\providecommand{\BIBforeignlanguage}[2]{{%
\expandafter\ifx\csname l@#1\endcsname\relax
\typeout{** WARNING: IEEEtran.bst: No hyphenation pattern has been}%
\typeout{** loaded for the language `#1'. Using the pattern for}%
\typeout{** the default language instead.}%
\else
\language=\csname l@#1\endcsname
\fi
#2}}
\providecommand{\BIBdecl}{\relax}
\BIBdecl

\bibitem{mao2018deep}
Q.~Mao, F.~Hu, and Q.~Hao, ``Deep learning for intelligent wireless networks: A
  comprehensive survey,'' \emph{IEEE Commun. Surv. Tutor.}, vol.~20, no.~4, pp.
  2595--2621, 2018.

\bibitem{liu2015joint}
J.~Liu, N.~B. Shroff, C.~H. Xia, and H.~D. Sherali, ``Joint congestion control
  and routing optimization: An efficient second-order distributed approach,''
  \emph{IEEE/ACM Trans. Netw.}, vol.~24, no.~3, pp. 1404--1420, 2015.

\bibitem{eisen2020optimal}
M.~Eisen and A.~Ribeiro, ``Optimal wireless resource allocation with random
  edge graph neural networks,'' \emph{IEEE Trans. Signal Process.}, vol.~68,
  pp. 2977--2991, 2020.

\bibitem{naderializadeh2023learning}
N.~NaderiAlizadeh, M.~Eisen, and A.~Ribeiro, ``Learning resilient radio
  resource management policies with graph neural networks,'' \emph{IEEE Trans.
  Signal Process.}, vol.~71, pp. 995--1009, 2023.

\bibitem{naderializadeh2022state}
------, ``State-augmented learnable algorithms for resource management in
  wireless networks,'' \emph{IEEE Trans. Signal Process.}, vol.~70, pp.
  5898--5912, 2022.

\bibitem{uslu2024learning}
Y.~B. Uslu, R.~Doostnejad, A.~Ribeiro, and N.~NaderiAlizadeh, ``Learning to
  slice wi-fi networks: A state-augmented primal-dual approach,'' \emph{arXiv
  preprint arXiv:2405.05748}, 2024.

\bibitem{xia2018utility}
S.~Xia, P.~Wang, and H.~M. Kwon, ``Utility-optimal wireless routing in the
  presence of heavy tails,'' \emph{IEEE Trans. Veh. Technol.}, vol.~68, no.~1,
  pp. 780--796, 2018.

\bibitem{zargham2013accelerated}
M.~Zargham, A.~Ribeiro, and A.~Jadbabaie, ``Accelerated backpressure
  algorithm,'' in \emph{Proc. IEEE Global Commun. Conf. (GLOBECOM)}, 2013, pp.
  2269--2275.

\bibitem{ribeiro2009stochastic}
A.~Ribeiro, ``Stochastic soft backpressure algorithms for routing and
  scheduling in wireless ad-hoc networks,'' in \emph{Proc. 3rd IEEE Int.
  Workshop Comput. Adv. Multi-Sensor Adaptive Process. (CAMSAP)}, 2009, pp.
  137--140.

\bibitem{xia2017stochastic}
S.~Xia and P.~Wang, ``Stochastic network utility maximization in the presence
  of heavy-tails,'' in \emph{Proc. IEEE Int. Conf. Commun. (ICC)}, 2017, pp.
  1--7.

\bibitem{xia2014distributed}
S.~Xia, ``Distributed throughput optimal scheduling for wireless networks,''
  Ph.D. dissertation, Wichita State University, 2014.

\bibitem{georgiadis2006resource}
L.~Georgiadis, M.~J. Neely, L.~Tassiulas \emph{et~al.}, ``Resource allocation
  and cross-layer control in wireless networks,'' \emph{Found. Trends Netw.},
  vol.~1, no.~1, pp. 1--144, 2006.

\bibitem{bernardez2023magnneto}
G.~Bern{\'a}rdez, J.~Su{\'a}rez-Varela, A.~L{\'o}pez, X.~Shi, S.~Xiao,
  X.~Cheng, P.~Barlet-Ros, and A.~Cabellos-Aparicio, ``Magnneto: A graph neural
  network-based multi-agent system for traffic engineering,'' \emph{IEEE Trans.
  Cogn. Commun. Netw.}, 2023.

\bibitem{valadarsky2017learning}
A.~Valadarsky, M.~Schapira, D.~Shahaf, and A.~Tamar, ``Learning to route,'' in
  \emph{Proc. 16th ACM Workshop Hot Topics Netw.}, 2017, pp. 185--191.

\bibitem{xu2018experience}
Z.~Xu, J.~Tang, J.~Meng, W.~Zhang, Y.~Wang, C.~H. Liu, and D.~Yang,
  ``Experience-driven networking: A deep reinforcement learning based
  approach,'' in \emph{Proc. IEEE Int. Conf. Comput. Commun. (INFOCOM)}, 2018,
  pp. 1871--1879.

\bibitem{geng2020multi}
N.~Geng, T.~Lan, V.~Aggarwal, Y.~Yang, and M.~Xu, ``A multi-agent reinforcement
  learning perspective on distributed traffic engineering,'' in \emph{Proc.
  IEEE Int. Conf. Netw. Protocols (ICNP)}, 2020, pp. 1--11.

\bibitem{sun2017learning}
H.~Sun, X.~Chen, Q.~Shi, M.~Hong, X.~Fu, and N.~D. Sidiropoulos, ``Learning to
  optimize: Training deep neural networks for wireless resource management,''
  in \emph{Proc. IEEE Int. Workshop Signal Process. Adv. Wireless Commun.
  (SPAWC)}, 2017, pp. 1--6.

\bibitem{lei2017deep}
L.~Lei, L.~You, G.~Dai, T.~X. Vu, D.~Yuan, and S.~Chatzinotas, ``A deep
  learning approach for optimizing content delivering in cache-enabled
  hetnet,'' in \emph{Proc. Int. Symp. Wireless Commun. Syst. (ISWCS)}.\hskip
  1em plus 0.5em minus 0.4em\relax IEEE, 2017, pp. 449--453.

\bibitem{xu2019energy}
D.~Xu, X.~Chen, C.~Wu, S.~Zhang, S.~Xu, and S.~Cao, ``Energy-efficient
  subchannel and power allocation for hetnets based on convolutional neural
  network,'' in \emph{Proc. IEEE Veh. Technol. Conf. (VTC-Spring)}, 2019, pp.
  1--5.

\bibitem{van2019sum}
T.~Van~Chien, E.~Bjornson, and E.~G. Larsson, ``Sum spectral efficiency
  maximization in massive mimo systems: Benefits from deep learning,'' in
  \emph{Proc. IEEE Int. Conf. Commun. (ICC)}, 2019, pp. 1--6.

\bibitem{de2018team}
P.~de~Kerret, D.~Gesbert, and M.~Filippone, ``Team deep neural networks for
  interference channels,'' in \emph{Proc. IEEE Int. Conf. Commun. Workshops
  (ICC Workshops)}, 2018, pp. 1--6.

\bibitem{meng2020power}
F.~Meng, P.~Chen, L.~Wu, and J.~Cheng, ``Power allocation in multi-user
  cellular networks: Deep reinforcement learning approaches,'' \emph{IEEE
  Trans. Wireless Commun.}, vol.~19, no.~10, pp. 6255--6267, 2020.

\bibitem{cui2019spatial}
W.~Cui, K.~Shen, and W.~Yu, ``Spatial deep learning for wireless scheduling,''
  \emph{IEEE J. Sel. Areas Commun.}, vol.~37, no.~6, pp. 1248--1261, 2019.

\bibitem{MEDHI201864}
D.~Medhi and K.~Ramasamy, ``Routing protocols: Framework and principles,'' in
  \emph{Network Routing}, 2nd~ed., ser. The Morgan Kaufmann Series in
  Networking, D.~Medhi and K.~Ramasamy, Eds.\hskip 1em plus 0.5em minus
  0.4em\relax Boston, MA, USA: Morgan Kaufmann, 2018, pp. 64--113.

\bibitem{jia2014caffe}
Y.~Jia, E.~Shelhamer, J.~Donahue, S.~Karayev, J.~Long, R.~Girshick,
  S.~Guadarrama, and T.~Darrell, ``Caffe: Convolutional architecture for fast
  feature embedding,'' in \emph{Proc. 22nd ACM Int. Conf. Multimedia}, 2014,
  pp. 675--678.

\bibitem{peng2017modulation}
S.~Peng, H.~Jiang, H.~Wang, H.~Alwageed, and Y.-D. Yao, ``Modulation
  classification using convolutional neural network based deep learning
  model,'' in \emph{Proc. 26th Wireless Opt. Commun. Conf. (WOCC)}, 2017, pp.
  1--5.

\bibitem{zhang2023admire}
J.~Zhang, Z.~Chen, B.~Zhang, R.~Wang, H.~Ma, and Y.~Ji, ``Admire: collaborative
  data-driven and model-driven intelligent routing engine for traffic grooming
  in multi-layer x-haul networks,'' \emph{J. Opt. Commun. Netw.}, vol.~15,
  no.~2, pp. A63--A73, 2023.

\bibitem{he2020machine}
J.~He, J.~Lee, S.~Kandeepan, and K.~Wang, ``Machine learning techniques in
  radio-over-fiber systems and networks,'' in \emph{Photonics}, vol.~7,
  no.~4.\hskip 1em plus 0.5em minus 0.4em\relax MDPI, 2020, p. 105.

\bibitem{shen2022graph}
Y.~Shen, J.~Zhang, S.~Song, and K.~B. Letaief, ``Graph neural networks for
  wireless communications: From theory to practice,'' \emph{IEEE Trans. on
  Wireless Commun.}, 2022.

\bibitem{henaff2015deep}
M.~Henaff, J.~Bruna, and Y.~LeCun, ``Deep convolutional networks on
  graph-structured data (2015),'' \emph{arXiv preprint arXiv:1506.05163}, 2015.

\bibitem{gama2018convolutional}
F.~Gama, A.~G. Marques, G.~Leus, and A.~Ribeiro, ``Convolutional neural network
  architectures for signals supported on graphs,'' \emph{IEEE Trans. Signal
  Process.}, vol.~67, no.~4, pp. 1034--1049, 2018.

\bibitem{solodova2024graph}
O.~Solodova, N.~Richardson, D.~Oktay, and R.~P. Adams, ``Graph neural networks
  gone hogwild,'' \emph{arXiv preprint arXiv:2407.00494}, 2024.

\bibitem{ribeiro2012optimal}
A.~Ribeiro, ``Optimal resource allocation in wireless communication and
  networking,'' \emph{EURASIP J. Wireless Commun. Netw.}, vol. 2012, no.~1, pp.
  1--19, 2012.

\bibitem{calvo2021state}
M.~Calvo-Fullana, S.~Paternain, L.~F. Chamon, and A.~Ribeiro, ``State augmented
  constrained reinforcement learning: Overcoming the limitations of learning
  with rewards,'' \emph{arXiv preprint arXiv:2102.11941}, 2021.

\bibitem{eisen2019learning}
M.~Eisen, C.~Zhang, L.~F. Chamon, D.~D. Lee, and A.~Ribeiro, ``Learning optimal
  resource allocations in wireless systems,'' \emph{IEEE Trans. Signal
  Process.}, vol.~67, no.~10, pp. 2775--2790, 2019.

\bibitem{nasir2019multi}
Y.~S. Nasir and D.~Guo, ``Multi-agent deep reinforcement learning for dynamic
  power allocation in wireless networks,'' \emph{IEEE J. Sel. Areas Commun.},
  vol.~37, no.~10, pp. 2239--2250, 2019.

\bibitem{liang2019deep}
L.~Liang, H.~Ye, G.~Yu, and G.~Y. Li, ``Deep-learning-based wireless resource
  allocation with application to vehicular networks,'' \emph{Proc. IEEE}, vol.
  108, no.~2, pp. 341--356, 2019.

\bibitem{liu2006cross}
J.~Liu, T.~Y. Park, Y.~T. Hou, Y.~Shi, and H.~D. Sherali, ``Cross-layer
  optimization on routing and power control of mimo ad hoc networks,'' Virginia
  Institute of Technology, Tech. Rep., 2006.

\bibitem{chatzipanagiotis2015augmented}
N.~Chatzipanagiotis, D.~Dentcheva, and M.~M. Zavlanos, ``An augmented
  lagrangian method for distributed optimization,'' \emph{Math. Program.}, vol.
  152, pp. 405--434, 2015.

\bibitem{bertsekas2015parallel}
D.~Bertsekas and J.~Tsitsiklis, \emph{Parallel and distributed computation:
  numerical methods}.\hskip 1em plus 0.5em minus 0.4em\relax Athena Scientific,
  2015.

\bibitem{ruszczynski2011nonlinear}
A.~Ruszczynski, \emph{Nonlinear Optimization}.\hskip 1em plus 0.5em minus
  0.4em\relax Princeton University Press, 2011.

\bibitem{chatzipanagiotis2017convergence}
N.~Chatzipanagiotis and M.~M. Zavlanos, ``On the convergence of a distributed
  augmented lagrangian method for nonconvex optimization,'' \emph{IEEE Trans.
  Autom. Control}, vol.~62, no.~9, pp. 4405--4420, 2017.

\bibitem{su2022distributed}
J.~Su, S.~Yu, B.~Li, and Y.~Ye, ``Distributed and collective intelligence for
  computation offloading in aerial edge networks,'' \emph{IEEE Trans. Intell.
  Transp. Syst.}, 2022.

\bibitem{boyd2011distributed}
S.~Boyd, N.~Parikh, E.~Chu, B.~Peleato, J.~Eckstein \emph{et~al.},
  ``Distributed optimization and statistical learning via the alternating
  direction method of multipliers,'' \emph{Found. Trends Mach. Learn.}, vol.~3,
  no.~1, pp. 1--122, 2011.

\bibitem{battaglia2018relational}
P.~W. Battaglia, J.~B. Hamrick, V.~Bapst, A.~Sanchez-Gonzalez, V.~Zambaldi,
  M.~Malinowski, A.~Tacchetti, D.~Raposo, A.~Santoro, R.~Faulkner
  \emph{et~al.}, ``Relational inductive biases, deep learning, and graph
  networks,'' \emph{arXiv preprint arXiv:1806.01261}, 2018.

\bibitem{wu2020comprehensive}
Z.~Wu, S.~Pan, F.~Chen, G.~Long, C.~Zhang, and S.~Y. Philip, ``A comprehensive
  survey on graph neural networks,'' \emph{IEEE Trans. Neural Netw. Learn.
  Syst.}, vol.~32, no.~1, pp. 4--24, 2020.

\bibitem{lee2020graph}
M.~Lee, G.~Yu, and G.~Y. Li, ``Graph embedding-based wireless link scheduling
  with few training samples,'' \emph{IEEE Trans. Wireless Commun.}, vol.~20,
  no.~4, pp. 2282--2294, 2020.

\bibitem{shen2019graph}
Y.~Shen, Y.~Shi, J.~Zhang, and K.~B. Letaief, ``A graph neural network approach
  for scalable wireless power control,'' in \emph{Proc. IEEE Globecom Workshops
  (GC Wkshps)}, 2019, pp. 1--6.

\bibitem{sandryhaila2014big}
A.~Sandryhaila and J.~M. Moura, ``Big data analysis with signal processing on
  graphs: Representation and processing of massive data sets with irregular
  structure,'' \emph{IEEE Signal Process. Mag.}, vol.~31, no.~5, pp. 80--90,
  2014.

\bibitem{knight2011internet}
S.~Knight, H.~X. Nguyen, N.~Falkner, R.~Bowden, and M.~Roughan, ``The internet
  topology zoo,'' \emph{IEEE J. Sel. Areas Commun.}, vol.~29, no.~9, pp.
  1765--1775, 2011.

\end{thebibliography}







\end{document}